\documentclass[10pt, conference, final, twocolumn]{IEEEtran}
\usepackage[a4paper,margin=1in]{geometry} 
\IEEEoverridecommandlockouts
\usepackage{macro}
\usepackage[normalem]{ulem}

\title{Efficient and rate-optimal list-decoding \\ in the presence of minimal feedback}
\date{}

\author{
  \IEEEauthorblockN{
    Pranav Joshi\IEEEauthorrefmark{1}, 
    Daniel McMorrow\IEEEauthorrefmark{2}, 
    Yihan Zhang\IEEEauthorrefmark{3}, 
    Amitalok Budkuley\IEEEauthorrefmark{4}, and 
    Sidharth Jaggi\IEEEauthorrefmark{3}
  }
  \IEEEauthorblockA{\IEEEauthorrefmark{1}California Institute of Technology, USA (pjoshi@caltech.edu)}
  \IEEEauthorblockA{\IEEEauthorrefmark{2}National University of Singapore, Singapore (dmcmorrow@u.nus.edu)}
  \IEEEauthorblockA{\IEEEauthorrefmark{3}University of Bristol, UK (\{yihan.zhang, sid.jaggi\}@bristol.ac.uk)}
  \IEEEauthorblockA{\IEEEauthorrefmark{4}Indian Institute of Technology Kharagpur, India (amitalok@ece.iitkgp.ac.in)}
  \thanks{This work was supported by the Engineering and Physical Sciences Research Council [Grant Ref: EP/Y028732/1]. A.~J.~Budkuley was partially supported by Dept. of Science and Technology, Govt. of India, through grant MTR/2023/001412.}
}

\begin{document}

\maketitle

\pagenumbering{gobble}  

\figuresatendfalse

\begin{abstract}
Given a channel with length-$n$ inputs and outputs over the alphabet $\{0,1,\ldots,q-1\}$, and of which a fraction $\varrho \in (0,1-1/q)$ of symbols 
can be arbitrarily corrupted by an adversary, a fundamental problem is that of communicating at rates close to the information-theoretically optimal values, while ensuring the receiver can infer that the transmitter's message is from a ``small" set.
While the existence of such codes is known, and constructions with computationally tractable encoding/decoding procedures are known for large $q$, we provide the first schemes that attain this performance for any $q \geq 2$, as long as low-rate feedback (asymptotically negligible relative to the number of transmissions) from the receiver to the transmitter is available. For any sufficiently small $\varepsilon > 0$ and $\varrho \in (1-{1}/{q}-\Theta(\sqrt{\varepsilon}))$ our minimal feedback scheme has the following parameters: 
Rate $1-H_q(\varrho) - \varepsilon$ (i.e., $\varepsilon$-close to information-theoretically optimal -- here $H_q(\varrho)$ is the $q$-ary entropy function), list-size $\exp\left(\mathcal{O}\left(\varepsilon^{-3/2}\log^2(1/\varepsilon)\right)\right)$, computational complexity of encoding/decoding $n^{\mathcal{O}(\varepsilon^{-1}\log(1/\varepsilon))}$, storage complexity $\mathcal{O}(n^{\eta+1}\log n)$ for a code design parameter $\eta>1$ that trades off storage complexity with the probability of error. The error probability is $\mathcal{O}(n^{-\eta})$, and the (vanishing) feedback rate is $\mathcal{O}({1}/{\sqrt{\log(n)}})$.
 
Our full-feedback scheme has zero probability of error and minimal storage complexity, while the other parameters are the same as the vanishing rate feedback scheme.

Our primary technique adapts
classic coding schemes in the information-theoretic literature for sources/channels with i.i.d.~statistics—the so-called \emph{Slepian-Wolf} and \emph{Weldon} schemes respectively—to the harder setting with adversarial noise.
\end{abstract}

\pagenumbering{arabic}  
\setcounter{page}{1}
\section{Introduction}
This work concerns itself with near-optimal schemes to reliably and computationally efficiently communicate a message at high rate in the presence of noise and some feedback. For such a fundamental problem there are many variants of interest in the literature:\\
\noindent $\bullet$ Random versus adversarial noise.\\
\noindent $\bullet$ Existential results with computational complexity scaling potentially exponentially in the length $n$ of the transmission, versus computationally efficient schemes.\footnote{In this work, in line with a wealth of literature in coding theory, we take a very lenient view of what counts as computationally efficient -- we will even consider schemes with complexity $n^{\cO(\poly(1/\varepsilon))}$ to be ``efficient", as long as $\varepsilon$ is a {\it fixed} (though possibly small) positive constant.}\\
\noindent $\bullet$ Specific alphabet sizes, versus arbitrary alphabet sizes.\\
\noindent $\bullet$ Schemes with positive but information-theoretically sub-optimal rates, versus those that approach these fundamental bounds.\\
\noindent $\bullet$ One-way communication, versus settings allowing (full or partial) feedback.\\
\noindent $\bullet$ Unique decoding (the decoder can precisely infer the transmitter's specific transmission), versus ``list-decoding" -- the goal of the receiver is to infer that the transmission belongs to a ``small" list.

Our interest in this work is in the latter flavour of each of the above settings -- the noise is adversarial, feedback (full or partial) from the receiver to the transmitter is allowed, and we present computationally efficient schemes for arbitrary alphabet sizes at rates approaching information-theoretic limits for the problem of list-decoding -- we {\underline{\bf fully resolve}} this setting. 

{\bf \underline{Channels with random noise}:}
Shannon's classic work~\cite{shannon1948mathematical} characterized  the optimal communication rate -- the {\it capacity} -- of discrete memoryless channels (DMCs) through a single-letter formula for the maximum communication rate achievable with vanishing error probability. 
A decade later Shannon \cite{shannon1956zero} then showed the counter-intuitive result that for DMCs, the feedback capacity is exactly equal to that without feedback. 
Nonetheless feedback may provide benefits in terms of code complexity or probability of error (e.g. \cite{li2015efficient}). Channels of specific interest are those with {\it symmetric noise} -- a symbol taking values in $\Brackets{q} \coloneqq \braces{0,1,\cdots,q-1}$ remains uncorrupted with probability $1-\varrho$, and with probability $\frac{\varrho}{q-1}$ gets corrupted into one of the remaining $q-1$ symbols -- for such channels the Shannon capacity is $1-H_q(\varrho)$. In particular, we highlight the {\it Horstein} scheme \cite{horstein1960sequential} for binary ($q=2$) symmetric channels; their generalization to general DMCs -- the so-called {\it posterior-matching} schemes \cite{shayevitz2011optimal,li2015efficient}, and  {\it variable-length block-coding schemes} -- the so-called {\it Weldon} schemes \cite{weldon1963asymptotic,ahlswede1971constructive,Ooi1998}. These have had significant impact in the information-theory literature of feedback communication, and the last class provides significant inspiration for our work. 
We also note in passing that in the random noise setting, due to ``strong converses" \cite{wolfowitz1957coding}, relaxing the unique decoding problem to list-decoding has no impact on the capacity, and that computationally efficient encoding/decoding schemes for general DMCs at rates approaching the Shannon capacity are known (e.g. \cite{forney1965concatenated, arikan2009channel}). Thus the random noise setting is quite well-understood.

{\bf \underline{Adversarial channels}:}
The capacity of channels with adversarial noise (and no feedback) is in general open. 
Shannon~\cite{shannon1956zero} showed that this capacity is determined by the maximum size of an independent set in the channel's {\it confusion graph}, but no general closed-form expression is known -- see \cite{lapidoth2002reliable, csiszar2011information, dey2024codes} for surveys on known results for worst-case noise models over general channels. Computing this capacity is challenging (\cite{lovasz1979}) -- indeed, it may well be non-computable (\cite{alon2006shannon}).

For ``simple" \emph{constrained} adversarial channels (that are of considerable interest in applications, and are the primary focus of this paper) where a fraction $\varrho$ of symbols may be arbitrarily corrupted by an adversary, when the alphabet-size is sufficiently large, say at least $n$, Reed-Solomon codes \cite{reed1960polynomial} meet the fundamental outer bounds on code sizes, the classic Singleton bound \cite{singleton1964maximum}. For codes with rates that are $\varepsilon$-close to the Singleton bound, constructions are known over alphabet sizes $q$ that are a large constant -- e.g. \cite{guruswami2005} works for  $q = \exp_2(\cO(\varepsilon^{-4}\log(1/\varepsilon)))$. But if the alphabet-size $q$ is a fixed {\it small} constant, the capacity is equivalent to the maximal sphere packing density in Hamming spaces, with the best upper and lower bounds on optimal rates differing significantly \cite{gilbert1952comparison, garcia1995tower, varshamov1957estimate,mceliece1977new, alrabiah2026binary, openai2026advances}.

{\bf \underline{Adversarial channels with feedback}:}
In the presence of full public\footnote{The situation is quite different if the feedback is private from the jammer -- see \cite{ahlswede1978elimination}.} noiseless feedback, it is known (see for instance \cite{berlekamp-thesis}) that the capacity can be strictly greater than without feedback (and even strictly greater than channels where the adversary has to design its jamming sequence as a causal function of the transmissions \cite{chen2015characterization} -- channels with feedback are implicitly causal, but this gap shows that feedback helps in addition to the help due to causality). 
For channels where a fraction $\varrho$ of symbols 
may be adversarially corrupted and full noiseless feedback is available, for the setting where the channel alphabet $\Brackets{q}$ is binary, a complete characterization of the capacity is known. In particular, the capacity equals the Shannon capacity $1-H(\varrho)$ for $\varrho<\frac{3-\sqrt{5}}{4}\eqqcolon\varrho^*$, is zero for $\varrho > 1/3$ (in contrast, in the binary setting with no feedback, no positive rate is achievable when $\varrho \geq 1/4$), and is a straight line interpolating between these two rates at $\varrho \in (\varrho^*,1/3)$ for intermediate values of $\varrho$. The converse for $\varrho\in[0,1/3]$ and achievability for the linear segment $\varrho\in[\varrho^*,1/3]$ was given by \cite{berlekamp-thesis}, and the achievability for $\varrho\leq\varrho^*$ provided by \cite{zigangirov-feedback}. On the other hand, for scenarios where the channel alphabet $q \geq 3$, a complete capacity characterization is not known. The converse, provided by \cite{ahlswede2006rubber}, similarly has two regimes, equaling $1-H_q(\varrho)$ for $\varrho<\nicefrac{1}{q}$, zero for $\varrho\geq1/2$, and is a straight line between these two curves for intermediate $\varrho$. For the linear portion, $\varrho\in[\nicefrac{1}{q},\nicefrac{1}{2}]$, the rubber method of \cite{ahlswede2006rubber} achieves capacity $(1-2\varrho)\log_q(q-1)$. 
For $\varrho<\nicefrac{1}{q}$, essentially the best-known achievability bounds come from the $r$-rubber generalization \cite{ahlswede2006rubber}, which produces a family of lines tangent to the converse curve and passing through the point $(\varrho, 0)$, with $\varrho$ taking the form $\varrho = 1/r$ for an arbitrary integer $r > q$. This construction specifies the capacity exactly at these countably many “contact points”, yet the full curve remains unknown.
A slight improvement to the achievability is given in \cite{lebedev2016rubber-improve}, but still does not match the converse. A concise survey of the rubber method and its variants is given in \cite{deppe2020rubber-review}. All of the previously mentioned feedback schemes are explicit streaming algorithms with constant per-symbol work (and hence $\cO(n)$ computational complexity overall). 
Recent work (for instance \cite{haeupler2015communication,gupta2023binary}) has focused on constructing efficient schemes for settings with full or partial feedback enabling communication against error fractions up to the levels when such communication at \emph{positive} rates is possible, but in general the rates attained by such schemes are far from information-theoretic outer bounds -- an excellent survey is available in \cite{gelles2017coding}.
Also, for $\poly(n)$-size alphabets  computationally efficient schemes matching information-theoretic outer bounds are known -- see for instance \cite{bakshi2025optimal}.

{\bf \underline{List-decoding for adversarial channels}:} A different variant of the communication problem, introduced in \cite{elias1957list, wozencraft1958list}, relaxes the requirement that the receiver be able to uniquely identify the transmitted message, and instead just requires that after communication (with adversarial noise), the receiver's uncertainty is over a ``relatively small" set -- say polynomial in the communication length $n$, or even a constant independent of $n$. If computational complexity of encoding/decoding is not a consideration, then the list-decoding capacity for general adversarial channels is well-understood at rates up to information-theoretic limits \cite{sarwate2012list}, with the $q$-ary bit flip channel (which is what we consider) having list-decoding capacity $1-H_q(p)$, where $p$ is the fraction of symbols the adversary can flip. However, adding the requirement of computational tractability in encoding/decoding makes the problem significantly more challenging.

Over sufficiently large alphabets (increasing as a function of $n$), 
the first polynomial time codes that achieved list-decoding capacity for large alphabets were given by Guruswami and Rudra in \cite{guruswami2007explicitcodesachievinglist}, who introduced Folded Reed-Solomon codes, and showed that such codes of rate $1-\varrho - \varepsilon$ can be decoded with list size $(n/\eps)^{\cO(1/\eps)}$, and runs in time $(n/\eps)^{\cO(1/\eps^2)}$.
Subsequent improvements on the list size were given by \cite{kopparty2018improved}, who showed that the bound on the worst case list size can be improved to $(1/\eps)^{\cO(1/\eps\log(1/\eps))}$. The results of \cite{guo2021efficient} show that codes over somewhat smaller alphabets of size $(1/\eps)^{\cO(1/\eps^2)}$ and rate $1-\varrho-\varepsilon$ can be decoded with list-size at most $\exp(\poly(1/\eps))$. Moreover, the codes can be encoded in time $\poly(1/\eps, n)$ and decoded in time $\exp(\poly(1/\eps))\poly(n)$. For smaller alphabets, including the important case of binary alphabets, it is unknown how to computationally tractably list-decode at rates close to the information theoretic limit of $1-H_q(\varrho)$ -- we are unaware of any codes that exceed the Zyablov/Blokh-Zyablov bounds -- a survey of some results can be found in \cite{guruswami2007algorithmic, guruswami2009list}.


{\bf \uline{List-decoding for adversarial channels with feedback}:} The literature on this variant, which is our primary focus, is sparse. As discussed above, there exists a ubiquitous upper bound on the rate of $1 - H_q(p)$ for $q$-ary bit flip channels, even when allowing full, noiseless feedback. In the binary alphabet setting with full feedback, the work of \cite{shayevitz2009} generalizes the techniques of Berlekamp \cite{berlekamp-thesis} to provide bounds on possible trade-offs between rates attainable and specific integer list-sizes -- it is unclear how to generalize these techniques to non-binary alphabets, or settings with only partial feedback available. Recent work in \cite{gupta2025list} examines the question again for binary alphabets and specific integer list-sizes, and the primary focus is on understanding thresholds of error fractions allowing for positive rates (rather than rates approaching the information-theoretic limits -- which are in any case poorly understood unless the list-size is allowed to be a large constant). There are also connections between this variant and the literature on searching with lies (Ulam's game) -- we refer the reader to the survey in \cite{pelc2002searching}.

\subsection{Our contributions and methods}
As the discussion above indicates, computationally efficiently list-decoding at information-theoretically near-optimal rates, especially over small alphabets, is challenging. Our primary contribution is to do so with an asymptotically negligible (in $n$) amount of feedback.
As a first step, in \Cref{thm:fullfb} we present a scheme that works in the presence of {\it full} feedback -- the transmitter observes, in a causal manner, all the symbols received by the receiver. Presenting this scheme as a precursor to our main result presented in \Cref{thm:partfb} (where only minimal feedback is available) aids exposition considerably, since the first scheme already contains many ideas in a conceptually simpler setting. Key ideas in our schemes are:

\noindent {\bf 1. \underline{Full feedback setting} (\Cref{thm:fullfb}, \Cref{sec:full-fb})}\\
\noindent {\bf (i) \underline{Weldon-type schemes for DMCs}:} In \hyperref[para:weldon_dmc]{Section 3.1.1} we describe the classic Weldon scheme. It is a multi-stage scheme, with the transmitter using each stage to transmit a compressed version of the error pattern introduced by the channel in the previous stages (recall the channel outputs are fully fed back to the transmitter). For instance, over a $q$-ary symmetric channel which corrupts symbols with probability $\varrho$, say in the first stage the transmitter transmits its $k$-symbol message with {\it no encoding}! The channel will corrupt roughly $k\varrho$ symbols, so in the second stage the transmitter uses roughly $kH_q(\varrho)$ symbols to describe this error pattern, again unencoded. Inductively, one can see that describing this error pattern in the $i$th stage requires about $k(H_q(\varrho))^{i-1}$ symbols. The computational complexity of encoding of each stage is low, ${\cO}(k)$, since it just requires compressing a sequence down to its entropy -- any off-the-shelf low-complexity scheme approaching information-theoretic compression rates (say arithmetic coding \cite{rissanen1979arithmetic}) may be used. After about $\cO(\log(k))$ stages the number of symbols that need to be transmitted is sublinear in $k$, and so to terminate, for the last stage even a scheme that is very information-theoretically inefficient (such as repetition coding with majority decoding) suffices, with the advantage that this has low computational complexity. Finally, the decoder reconstructs the noise-free transmission in each stage in reverse order (again with low computational complexity of decoding, since it is just decompression of a compressed sequence) to finally decode the first stage, which is just the message that the transmitter wishes to communicate. One can observe that the amount of communication required scales geometrically in the number of stages, requiring overall about $\nicefrac{k}{(1-H_q(\varrho))}$ symbols to be transmitted, hence the rate of this scheme approaches information-theoretic limits.

\noindent {\bf (ii) \underline{Handling non-uniformity of noise levels}:} If the noise is adversarial instead of random, several challenges arise in instantiating a Weldon-type scheme, the first of which is that for any $i$,  the fraction $p_i$ of symbols corrupted in the $i$th stage need not be close to $\varrho$. (Notationally we use $p_i$ to denote the noise fraction in the $i$th stage, to distinguish it from $\varrho$, the overall fraction.) In this case the number of symbols $k_i$ transmitted in the $i$th stage would be about $k_i = k\prod_{j=1}^{i-1}H_q(p_j)$. One then needs to argue that {\it regardless} of the adversary's choice of noise levels $\{p_i\}$, the overall number of symbols $\sum_{i}k_i $ that need to be transmitted in a Weldon-type scheme is never more than $\nicefrac{k}{(1-H_q(\varrho))}$. While arguing this directly seems challenging, it follows naturally from the convexity (in $\varrho$) of mutual information -- see \Cref{sec:app:descent} and \hyperref[para:weldon-adv]{Section 3.1.2}. Arguably the analysis in this section is the technical heart of this work.

\noindent {\bf (iii) \uline{Termination scheme}:} \phantomsection\label{para:fullfb-term}Another challenge with adapting the Weldon scheme to adversarial noise is that the final stage also needs to be resilient to adversarial noise. Our analysis in \hyperref[para:term_scheme]{Section 3.1.3} and \Cref{sec:app:descent} shows that if the transmitter's overall transmission rate is at least $\eps$ below the information-theoretic limit of $1-H_q(\varrho)$, regardless of the adversary's choice of noise levels $\{p_i\}$, after $\tilde{\lambda} = \cO(\eps^{-3/2}\log(1/\eps))$ (i.e., a constant, albeit a large constant) number of stages the residual fraction of noise available to the adversary is sufficiently small (bounded away from $1-\nicefrac{1}{q}$) enough for an off-the-shelf computationally efficient $q$-ary list-decoding scheme (adapted from~\cite[Theorem 3.1]{kopparty2018improved})) may be used by the receiver to list-decode the last-stage. The rate of this scheme is far from the information-theoretic limit of $1-H_q(\varrho)$, but since (as also shown in \Cref{sec:app:descent}) the residual entropy that needs to be communicated in this last stage is small, this existing scheme suffices.\\
\noindent {\bf (iv) \underline{Synchronization}:} \phantomsection\label{para:sync} The last issue that we need to handle in this adversarial setting is that the receiver in general may not know the noise levels $\{p_i\}$ across stages. Since in Weldon-type schemes the length of the $i$th stage depends on the noise-levels of stages preceding it, this could cause issues of synchronization, since the receiver may not be able to infer which contiguous subsequences of its observations correspond to which stage.
Here we make the observation that since our goal is just to list-decode, in principle the receiver can just guess all possible values of each of the $p_i$'s (suitably quantized, so that there's a constant number of possible $p_i$ for any stage $i$ with length $\ell_i$ symbols, rather than potentially $\ell_i+1$ values from the set $\{0,\nicefrac{1}{\ell_i},\ldots,\nicefrac{(\ell_i-1)}{\ell_i},1\}$). Combining with the analysis in \Cref{sec:app:descent} showing that a constant $\tilde{\lambda}$ number of Weldon-type stages suffice to ensure that the termination scheme is triggered, implies that the receiver only needs to make a constant number of guesses for this set $\{p_i\}$, each of which is itself concomitant with a constant number of guesses for the termination scheme, and therefore an overall constant number of guesses for the message. All of the receiver's potentially incorrect guesses are hence accounted for in the list size.
%
%
%
\\ \noindent {\bf 2. \uline{Asymptotically negligible feedback setting} (\Cref{thm:partfb}, \Cref{sec:partial_fb}})
In our model we allow $\cO(\nicefrac{1}{\sqrt{\log(n)}})$-rate feedback, i.e., after every $\kappa n$ symbols (for suitably small positive constant $\kappa$) sent from the transmitter to the receiver over the adversarially controlled channel, the receiver can send back $\kappa\cO(\nicefrac{n}{\sqrt{\log(n)}})$ symbols back noiselessly and publicly -- hence the feedback is asymptotically negligible in $n$. This feedback is also observed by the adversarial jammer, which can then base its future jamming patterns as a function of these observations and prior transmissions from the transmitter to the receiver\footnote{\label{fn:fix-so} As specified below, our model dictates a fixed speaking order {\it a priori} -- after every $\kappa n$ symbols transmitted forward, there are $\kappa\cO(\nicefrac{n}{\sqrt{\log(n)}})$ symbols transmitted back. As noted in~\cite{haeupler2014}, subtle issues may arise when the speaking order itself may be adaptive and dependent on what each party observes -- indeed, the speaking order itself, and even the decision to speak or not, may carry information. Our model choice is designed to prevent such arguably undesirable artifacts.}.
\\
\noindent {\bf (i) \uline{Slepian-Wolf/Weldon schemes with partial feedback, for random noise}:}
We first revisit Weldon-type schemes for channels with random noise in this setting.
In this setting clearly the transmitter can have at best imperfect knowledge of the error pattern induced by the channel, and so, {\it prima facie}, a Weldon-type scheme would not work. However, the classic distributed coding scheme of Slepian and Wolf \cite{slepian2003noiseless} offers an alternative. The idea, as outlined in \hyperref[para:sw]{Section 4.1.1}, is that if $\vbfx$ and $\vbfy$ are length-$n$ vectors drawn i.i.d. from a joint distribution $P_{\bfx,\bfy}$, then sufficiently many random hashes of $\vbfx$ (about $nH(\bfx|\bfy)$) along with $\vbfy$ suffice to reconstruct $\vbfx$ with high probability. That is, the vector $\vbfx$ can be compressed down to the information-theoretic limit of $nH(\bfx|\bfy)$ (which can never be larger and is in general smaller than $nH(\bfx)$, the entropy of $\vbfx$ itself) {\it despite} the encoder of $\vbfx$ having no knowledge of the vector $\vbfy$ other than the joint statistics of $\vbfx$ and $\vbfy$. In the Weldon-type setting $\vbfx$ represents the transmitter's transmission at stage $i$, $\vbfy$ represents the channel output of stage $i$, and the hashes represent the transmission of stage $i+1$. While the compression scheme in \cite{slepian2003noiseless} is computationally inefficient, via the ``usual" trick of concatenation (i.e., as outlined in \hyperref[para:sw-concat]{Section 4.1.2}, breaking each stage down into small chunks of length $\cO(\log(n))$, and doing brute-force encoding/decoding on these chunks, with an outer code to clean up any residual errors), the scheme can be made computationally tractable.

\noindent \phantomsection\label{para:ii}{\bf (ii) \uline{Adversarial setting: noise-level estimation with partial feedback}:} The first challenge to overcome in instantiating the above partial-feedback Slepian-Wolf/Weldon-type scheme in adversarial settings is to ensure that the transmitter has a reasonable estimate of the joint statistics of $\vbfx$ and $\vbfy$ for each stage, especially in the face of an adversary which dynamically varies its noise levels. To do so, as outlined in \hyperref[para:est-nl]{Section 4.2.1}, the receiver periodically samples a small random subset of its observed channel outputs, and uses the noiseless partial feedback channel to describe to the transmitter the indices and observed values of these channel outputs. As described in \Cref{sec:est-noise} this enables the transmitter to come up with a ``good enough" estimate of the noise level in any given stage.

\noindent \phantomsection\label{para:iii}{\bf (iii) \uline{Adversarial setting: ``quasi-uniformizing" noise in each chunk}:} The next issue to resolve is that for the concatenated coding scheme within each stage $i$ to operate at rates approaching information-theoretic limits, at a minimum the noise-level within each length-$\cO(\log(n))$ chunk of stage $i$ should be ``close" to the overall noise-level across the entire stage. The adversary of course has no incentive to ensure this. To deal with this non-uniformity, we use another idea extant in the literature (see for instance \cite{ahlswede1986arbitrarily,guruswami2016optimal}) -- in each stage, the transmitter scrambles the indices of the transmitted symbol using a permutation selected uniformly at random from a polynomial-sized set. Information about which permutation was used is only sent to the receiver in the {\it next} stage, by when it is too late for it to be useful to the adversary in choosing its noise pattern. In particular we show (see \hyperref[para:perm]{Section 4.2.2} and \Cref{sec:perm}) that, regardless of the adversary's choice of noise pattern, with high probability over the randomness in the scrambling permutation used by the transmitter, the noise level in most chunks in each stage is close to the overall average noise level (and the small fraction of chunks for which it is not can be dealt with via the outer code of our concatenation scheme).

\noindent \phantomsection\label{para:iv}{\bf (iv) \uline{Adversarial setting: adversarially robust hashes}:} Even after ensuring that most chunks of our concatenated coding scheme have roughly the same ``quasi-uniform" noise level, we need to argue that our random hashes are resilient to adversarially chosen ``quasi-uniform" noise patterns as well. To do so, as we show in \hyperref[para:hash]{Section 4.2.3}, it suffices to add another layer of randomness, by requiring the transmitter to choose randomly from one of $q^{\Theta(\sqrt{\log(n)})}$ hash-tables -- the choice of which hash-table to use is coordinated via a random $\Theta(\sqrt{\log(n)})$-length seed sent back by the receiver. We show that for any quasi-uniform noise pattern injected by the adversary, for a vast majority of these seeds our concatenation scheme nonetheless works.

\noindent {\bf (v) \underline{Adversarial setting: synchronization}:} Finally, in \Cref{subsec:partialfb-sync,subsec:partfb-recursion}, we collate the ideas above into a Weldon-type scheme for adversarial models in the presence of partial feedback. In particular, a couple of protocol-level synchronization issues pop up regarding the receiver's lack of knowledge of noise-levels, and him not knowing when to send relevant feedback. Padding our scheme with appropriate amounts of padding/slack resolves these issues. This Section also deals with careful choice of internal parameters for the scheme 
to ensure the declared performance.

We note that our scheme for the full feedback is completely deterministic, and thus succeeds with probability $1$. On the other hand, our scheme for the partial feedback setting involves using internal randomness in selecting the subsets of channel outputs \hyperref[para:ii]{Paragraph (ii)}, the choice of permutation in each chunk \hyperref[para:iii]{Paragraph (iii)}, and the choice of hash tables \hyperref[para:iv]{Paragraph (iv)}. Therefore, the scheme succeeds with probability $1-o(1)$ over this internal randomness, rather than with probability $1$ as in the full feedback case. Unlike a random codebook, whose description and decoding may require exponential time complexity, all randomness in our construction generates polynomial-size objects. We leave for future work as to whether there exists a deterministic algorithm that succeeds with probability $1$ in the partial feedback setting.
\section{Model/Our Results}
\label{sec:model}

\paragraph{Notational conventions}
We denote random variables by lower case boldface letters, their values by lower case letters, and sets 
by calligraphic letters (e.g., $\bfx$, $x$ and $\cX$ respectively). Random vectors and their instantiations are denoted similarly with an under-bar (e.g. $\vbfx$, $\vx$), with elements indexed by subscripts (e.g., $\bfx_i, x_i$). We use $|\vx|$ to denote the length of a vector --  when not specified/it causes no confusion, vectors will usually be of length $n$.
Given vector $\vx=(x_1, x_2, ... , x_n)$ and indices $1\leq i\leq j\leq n$, we denote the contiguous sub-vector $\vx_{i:j}\coloneqq (x_i, x_{i+1}, ... , x_j)$. More generally, for any index set $S\subset[n]$, we write $\vx_S\coloneqq (x_t)_{t\in S}$.
Let $\cB(\vx,r)$ denote a Hamming ball centered at $\vx$ of radius $r$.
Let $[a]:=\{1,2,\cdots, a\},$ and $ \Brackets{b} \coloneqq \braces*{0,1,\cdots,b-1} $ where $a\in\mathbb{Z}_{\ge0},b\in\mathbb{Z}_{\ge1}$. For any set $\mathcal{A}\subseteq \cX,$ let 
$\ind{\cA}(x)$ denote the indicator of event $\{x\in\cA\}.$
The {\it type} (empirical p.m.f. of the elements) of $\vx$ is denoted $T_{\vx}$, and for a p.m.f. $P_\bfx$ on $\cX$ we define the {\it $\eps$-typical set} (set of all vectors close to a given type) as $\cT^n_\eps(\bfx) = \{\vx\in\cX^n : \lVert T_{\vx} - P_\bfx \rVert_{\infty} \leq \eps \}$, where $\lVert\cdot\rVert_\infty$ denotes the $\ell_\infty$ norm on $\bR^{|\cX|}$. For $q \geq 2$ we denote  $H_q(\cdot)$ and $H_q^{-1}(\cdot)$ as the $q$-ary and inverse $q$-ary entropy functions respectively.
Let $\wth{\vx}$ and $\disth{\vx}{\vy}$ respectively denote the Hamming weight of $\vx\in\cX^n$ and the Hamming distance between $\vx,\vy\in\cX^n$. 

\paragraph{Problem setup:}
We consider the problem of one-way communication from a transmitter Alice to a receiver Bob through a $q$-ary (for a fixed integer $q\ge2$) symbol flip channel controlled by a \emph{causal} (a.k.a.\ \emph{online}) adversary James while allowing (full or partial) instant noiseless feedback from Bob back to Alice. 
Specifically, let $\ul{m}\in\Brackets{q}^k$ be an arbitrary $q$-ary message of length $k$ that Alice wishes to transmit to Bob. 
Instead of transmitting $\ul{m}$ per se, Alice transmits encoded symbols $ x_i $ sequentially for each $i=1,2,\cdots,n$. 
The $i$th encoded symbol $x_i$ is a function of the message $\ul{m}$ and the noiseless feedback $ (f_1, \cdots, f_{i-1}) $ from Bob up to the previous step (in particular, for $i=1$, no such feedback is available). 
At time $i$, James, on observing the  symbols $\vx_{1:i}$ transmitted thus far  and the cumulative feedback $ (f_1, \cdots, f_{i-1}) $, 
chooses an error symbol $ s_i\in\Brackets{q} $ based on $\vx_{1:i}, \vf_{1:i-1}$. 
The channel then outputs the corrupted symbol $ y_i = x_i \oplus s_i \coloneqq x_i + s_i \bmod q $. 
Bob receives $y_i$ and sends a feedback sequence $ f_i \in \Brackets{q}^{b_i} $ of length $b_i\ge0$ (the case of $b_i = 0$ means no feedback) based on his received symbols thus far $ \vy_{1:i} $. We will be particularly interested in schemes with {\it fixed speaking order} (see Footnote \ref{fn:fix-so}), where the set of indices with $b_i \neq 0$ is determined in advance of communication and known to all parties.
The feedback sequence $f_i$ is instantly, noiselessly and publicly available to both Alice and James. Alice then proceeds to prepare the $(i+1)$th transmission. 
Finally, Bob, as a function of received symbols $ \vy \coloneqq (y_1, \cdots, y_n) $, outputs a list $\cL\subset\Brackets{q}^k$ of messages which must contain Alice's message $\ul{m}$. 

James's corruption is subject to a global constraint that at most a $\varrho$ fraction (for some $ \varrho \in (0,1) $ fixed) of transmitted symbols can be corrupted, i.e., $ \wth{\vs} \le n\varrho $ where $ \vs \coloneqq (s_1, \cdots, s_n) $. 
We call the collection of maps that Alice uses to generate $ \vx \coloneqq (x_1, \cdots, x_n) $ the encoding function $ \enc $, the collection of maps that Bob uses to generate $ (f_i)_{i = 1}^{n-1} $ the feedback function $ \mathrm{FB} $,\footnote{Note that $f_n$ in response to the $n$th transmitted symbol $x_n$ of Alice is not needed.} and the map that Bob uses to generate $\cL$ the decoding function $\dec$. 
The tuple of functions $ (\enc,\dec,\mathrm{FB}) $ constitutes a coding scheme. 
We allow Alice and Bob to locally randomize their encoding and decoding/feedback functions, respectively, using randomness that is only known to themselves and independent of other randomness in the system. 
\textbf{In particular, no shared randomness is allowed.} 
The functions $\enc,\dec,\mathrm{FB}$ are known to all parties before communication takes place. 
However, the realizations of $\ul{m}$ and the local randomness (if any) of any party are not revealed globally.\footnote{It actually transpires that our main results, \Cref{thm:fullfb,thm:partfb} hold even if James knows $\ul{m}$ in advance!} 

We say that full feedback is available if Bob simply sends back his observed symbols to Alice, i.e., $ (f_1, \cdots, f_{n-1}) = (y_1, \cdots, y_{n-1}) $. 
This immediately implies that Alice can infer the previous error symbols $ \vs_{1:i-1} $ at any time $i$. 
We say that only partial feedback is available if $ R_{\mathrm{FB}} < 1 $ -- this does not allow Bob to feedback all of $\vy$ -- he has to transmit some (possibly randomized) function of it. 

For a fixed alphabet size $q\ge2$ and an error budget $\varrho\in(0,1)$, there are three fundamental parameters of interest: the rate $ R \coloneqq \frac{k}{n} $ of communication from Alice to Bob, the rate $ R_{\mathrm{FB}} \coloneqq (n-1)^{-1} \sum_{i = 1}^{n-1} b_i $ of feedback from Bob to Alice, and the size $L \coloneqq \abs{\cL}$ of Bob's output list. 
From a computational perspective, the encoding/decoding/feedback functions are desired to be computationally efficient, i.e., with run time and construction time both polynomial in $n$. 
The main results of this work (Theorems~\ref{thm:fullfb},\ref{thm:partfb}) offer such communication schemes that operates at a rate $R$ arbitrarily close to the information-theoretic limit $ 1 - H_q(\varrho) $ for any $ \varrho < 1 - 1/q $ (so that $ 1 - H_q(\varrho) > 0 $) and outputs a list of constant (independent of $n$) size, with the aid of vanishing rate feedback, i.e., $ R_{\mathrm{FB}} = o(1) $.

\begin{theorem}[Full feedback] \label{thm:fullfb}
    For any $q\geq 2$, sufficiently small $\eps > 0$, let $\varrho \in \parens*{0, 1-\frac{1}{q}-\Theta\parens*{\sqrt{\eps}} }$ be James' budget in a $q$-ary symbol error channel with full feedback. 
    For sufficiently large $n$, there exists a coding scheme of rate $R = 1 - H_q(\varrho) - \eps$ with construction and decoding time each $n^{\cO(\eps^{-1}\log(1/\eps))}$
%
%
    \noindent such that for every message $\vm \in \alphX^{nR}$ with probability $1$ the decoder outputs a list $\cL$ of size $\exp{\parens*{\cO\parens*{\eps^{-3/2}\log^2(1/\eps)}}}$ containing $\vm$.
%
\end{theorem}

%
%
\begin{theorem}[Minimal feedback] \label{thm:partfb}
    For any $q\geq 2$, sufficiently small constant $\eps > 0$, let \newline $\varrho \in (0, 1-\frac{1}{q}-\Theta\parens{\sqrt{\eps}})$ be James' budget in a $q$-ary symbol error channel with $\cO(\nicefrac{1}{\sqrt{\log(n)}})$-rate feedback. 
%
For any $\eta > 1$ and sufficiently large $n$, there exists a coding scheme of rate $R = 1 - H_q(\varrho) - \eps$ with construction and decoding time each $n^{\cO(\eps^{-1}\log(1/\eps))}$, and storage complexity $\cO(n^{\eta+1} \log n)$
%
    \noindent such that for every message $\vm \in \alphX^{nR}$ with probability $1 - \cO(n^{-\eta})$ the decoder outputs a list $\cL$ of size $\exp{\parens*{\cO\parens*{\eps^{-3/2}\log^2(1/\eps)}}}$ containing $\vm$.
%
\end{theorem}

\section{Full feedback setting (\Cref{thm:fullfb})}
\label{sec:full-fb}
%
%
\subsection{Weldon-type schemes}
\label{subsec:weldon}
A Weldon-type scheme proceeds in \emph{stages}, $\vbfx = \parens*{\stgone, \stgtwo, ..., \vbfx^{(\nstages)}, \vbfx^{(\nstages+1)}}$, having, in general, varying lengths $\ell_i\coloneqq |\stgi|$. The entire unencoded message is sent as stage 1, $\stgone = \vbfm \in \alphX^{nR},\; R=1-H_q(\varrho)-\eps$. Bob receives the corrupted $\vbfy^{(1)} = \stgone\oplus \vbfs^{(1)}$. Alice receives $\ystgone$ as feedback and can hence deduce $\vbfs^{(1)}$. The basic idea of these schemes is that Alice's following stage, (here, $\stgtwo$), is a compressed version of the error pattern in the previous stage.

{\bf (3.1.1) For $q$-ary symmetric channels ($q$-SCs).}\phantomsection \label{para:weldon_dmc}
The $q$-SC$(\varrho)$ corrupts $\approx\ell_1 \varrho = nR\varrho$ symbols in $\stgone$. There are thus $\approx \binom{\ell_1}{\ell_1 \varrho}$ possible error patterns that could have occurred. Alice sends the index of this error pattern using $\log\bracks*{\binom{\ell_1}{\ell_1 \varrho}} \approx \ell_1H_q(\varrho)$ symbols as the next stage's transmission, $\stgtwo \in \Brackets{q}^{\ell_1H_q(\varrho)}$. She repeats this process for the following stages. We then have
    $\ell_{i+1} \approx \ell_i \cdot (H_q(\varrho))$ which implies that $\ell_i \approx nR(H_q(\varrho))^{i-1}$.
When the length of the last stage is sublinear in $n$, we execute a low-rate \textbf{termination scheme} - in the case of a $q$-SC, a simple repetition code suffices. Alice prepares the index of the error pattern of the last stage, ${\vbfz}^{(\lambda)}$. She sends $\vbfx^{(\lambda+1)}$ via an $r$-repetition code of $\vbfz^{(\lambda)}$. The scheme is decoded in reverse - Bob decodes $\vbfy^{(\lambda+1)}$, i.e. the repetition stage, by majority vote. Using $\vbfy^{(\lambda)}$ and the estimate of the error index $\vbfz^{(\lambda)}$, Bob obtains $\vbfx^{(\lambda)}$.  Recursively rolling back, Bob decodes Alice's original message $\vbfm = \stgone$. It can be shown that for $\lambda = \log\parens*{\frac{n}{\log(n)}}$ and $r=\log^2(n)$, the probability of error tends to zero (see for example \cite[Sec. 17.1.3]{elgamal2011network}).

{\bf (3.1.2) For adversarial channels.} \phantomsection\label{para:weldon-adv}
The adversarial setting differs in that James can attack each stage $i$ with a different fractional amount $p_i$. Again $\stgone\coloneq \vbfm$. The recursive algorithm becomes: Alice sends $\stgi\in \alphX^{\ell_i}$. James attacks a fraction $p_{i}$ of the symbols in $\stgi$ and Bob receives the noisy $\ystgi \in \alphX^{\ell_i}$. Alice transmits the \emph{index} of the error pattern induced by James, $\stgii \in \alphX^{\ell_i H_{q}(p_{i})}$ across the channel. This makes the length $\ell_i$  of the $i$th stage approximately $ \ell_{i-1}\cdot H_q(p_{i-1}) \approx nR\prod_{j=1}^{i-1} H_q(p_j)$ (see \Cref{fig:weldon-intro}).

%

\insertfig{fig:weldon-intro}{
    The progression of the Weldon scheme for adversarial channels. James corrupts a fraction $p_i$ of transmissions in stage $i$, resulting in the length of the next stage $\ell_{i+1} \approx \ell_i H_q(p_i)$. The termination stage occurs at the end.
}{
    \includegraphics[trim=45mm 170mm 30mm 55mm, clip, width=0.99\linewidth]{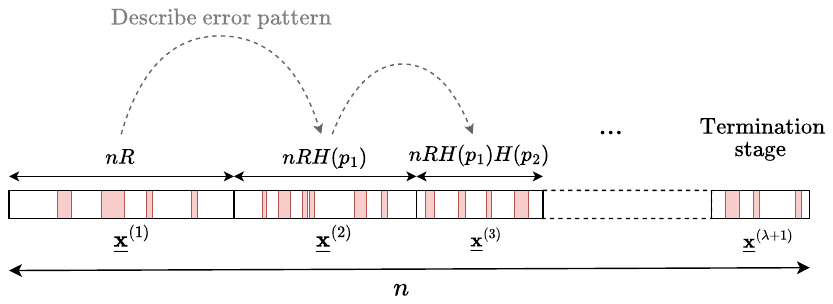}
}{
    Weldon-type schemes for adversarial channels (\hyperref[para:weldon-adv]{Section 3.1.2})
}

To analyse its performance, we introduce the following notation, visually depicted in \Cref{fig:weldon-fullfb-notation}. The length of stage $i$ is denoted $\ell_i$, the remaining symbols left before the scheme terminates is denoted $n_i$, and the fraction of errors James introduces is $p_i$ (i.e. $\ell_ip_i$ symbols are in error). The ``residual" of any quantity refers to the fractional amount of that quantity with respect to the remaining (residual) blocklength $n_i$. 
For example, the residual rate $R_i = \nicefrac{\ell_i}{n_i}$: the remaining ``message" length we have to transmit as a fraction of the remaining codeword length.
James' residual budget $\varrho_i$ is the number of channel errors he has left to introduce (at the \emph{start} of stage $i$), as a fraction of $n_i$, i.e. $\varrho_i = \frac{1}{n_i}\parens*{n_1\varrho_1 - \sum_{j=1}^{i-1}\ell_j p_j}$.
It is useful to derive the following recurrence relationships between the residual quantities for all $ i\in \{1, 2, ..., \nstages-1\}$. We have $n_{i+1} = n_i - \ell_i$, initial conditions: $\varrho_1 = \varrho, R_1 = R$ (\Cref{fig:weldon-fullfb-notation}), and
%
%

\begin{align} \label{eq:clean-recurs-ell-update}
    R_{i+1} &= \frac{\ell_{i+1}}{n_{i+1}} = \frac{\ell_iH_q(p_i)}{n_i-\ell_i} \overset{(a)}{=} \frac{R_iH_q(p_i)}{1-R_i}
    \\
    \varrho_{i+1} &\overset{(b)}{=} \frac{n_i\varrho_i - \ell_ip_{i}}{n_{i+1}} \overset{(c)}{=} \frac{\varrho_i - R_ip_i}{1-R_i} \nonumber,
\end{align}
where $(a),(c)$ follow from dividing throughout by $n_i$ and $(b)$ follows from noting that the remaining budget at stage $i$ was $n_i\varrho_i$ and the budget spent on stage $i$ was $\ell_ip_i$. We also note that the sum of the stage lengths $\ell_i$ up to and including the termination stage do not exceed the blocklength $n$, so there is no ``overflow" in this regard -- see \Cref{subsec:app:clean-descent} for details.
%
%

{\bf (3.1.3) Termination scheme.} \phantomsection \label{para:term_scheme} 
For our termination scheme, we use a concatenated code with a random inner code and a {\bf Folded Reed-Solomon} outer code  (see e.g., \cite[Sec. 17.1]{guruswami2012essential}). For Folded Reed-Solomon codes \cite{kopparty2018improved} showed the following list-recovery guarantee 
(see Theorem 3.1 and Section 1.2 in \cite{kopparty2018improved}):
\begin{lemma}
    \label{lem:FRS} \emph{(Adapted from \cite[Theorem 3.1]{kopparty2018improved})} Let $R \in (0,1)$, $\eps \in (0,1)$, and $\ell \in \bN$ be constants, and let $s  \in \bN$ be a prime power. Then, a Folded Reed-Solomon code with rate $R$ over an alphabet of size $s^{\cO(\ell/\gamma^2)}$ is $(1-R-\gamma, \ell, L)$-list-recoverable with $L = (\ell/\gamma)^{\cO(\frac{1}{\gamma}\log (\ell/\gamma))}$.
\end{lemma}
For our inner codes, we use codes of rate $r \in (0,1)$ and $(H_q^{-1}(1-r - \Theta(1/\ell), \ell)$ list-decode each chunk with the choice of $\ell = \Theta(1/\gamma)$. Combining this with the guarantee in \Cref{lem:FRS}, it then follows that the concatenated code has a rate of $r R$ and is $\big((1-R-\gamma)H_q^{-1}(1-r-\gamma), (1/\gamma^2)^{\cO(\frac{1}{\gamma}\log(1/\gamma))}\big)$ list-decodable. Moreover, construction and decoding can be done in $n^{\cO(\gamma^{-2}\log(1/\gamma))}$ time (see \cite[Remark 5.2]{guruswami2007explicitcodesachievinglist}). Optimizing over the values of $r$ and $R$, we can decode up to rates arbitrarily close to the {\bf Zyablov Bound} \cite[Sec. 10.2]{guruswami2012essential}, which for a decoding radius $\varrho \in (0, 1 - \frac{1}{q})$ and parameter $\eps_Z \in (0,1)$ equals $R_{\rm Z}(\varrho) =\underset{0 < r < 1 - H_q(\varrho + \eps_Z)}{\max} r\left(1 - \frac{\varrho}{H_q^{-1}(1 - r) - \eps_Z}\right)$.
Therefore, a sufficient condition to enter the termination scheme is for the residual rate $R_i$ to lie below the Zyablov bound $R_Z$, since any rate below this is achievable by our choice of termination scheme. We demonstrate in \Cref{subsec:app:clean-descent} that at most $\lambda =\cO(\gamma^{-3}\log(1/\eps))$ stages are required to reach this termination condition, where $\gamma$ is the value such that James' residual budget is $1 - \frac{1}{q} - \gamma$ in the termination scheme, by providing a lower bound on the reduction in residual rate in each stage.

\subsection{Synchronization}
\label{subsec:fullfb-sync}
As described in \hyperref[para:sync]{Paragraph (iv)}, in the adversarial setting, James' attack fraction in stage $i$, $p_i$, is known only to Alice (since she has access to both $\vbfx$ and $\vbfy$), but not Bob. Therefore, Bob does not know where each stage begins and ends (except stage 1).
To address this, Bob brute-force guesses all possible error fractions $\hat{\vp} = (\hp_1, \hp_2,..., \hp_{\nstages})$ in a discretized fashion, i.e. $\hp_i \in \{0, \delta, 2\delta, \dots 1 - \delta, 1\}$. From \hyperref[para:term_scheme]{Section 3.1.3}, for a particular guess $\hat{\vp}$, Bob obtains a worst case list size of $(1/\gamma^2)^{\cO(\frac{1}{\gamma}\log(1/\gamma))}$ due to \Cref{lem:FRS}; so after making all $(1/\delta)^{\nstages}$ possible guesses, $|\cL| \leq (1/\delta)^{\nstages} \cdot (1/\gamma^2)^{\cO(\gamma^{-1}\log(1/\gamma))}$. To maintain synchronization, at every stage $i$, Alice also rounds to the $\delta$ grid -- she sends the next state with length $\ell_{i+1} = \ell_i\cdot H_q(\hp_i)$, where $\hp_i = \ceil{p_i/\delta}\cdot\delta$ is a rounding \emph{up}. Since $\hp_i-p\leq\delta$, this incurs a small rate penalty at each stage; to ensure we still terminate succesfully, we choose $\delta = \Theta\parens*{\eps^{5/2}\log^{-2}(1/\eps)}$ (see \Cref{subsec:fullfb-imperfect-recursion}) and show that $\gamma = \Omega(\sqrt{\eps})$ (see \Cref{subsec:app:clean-descent}). Plugging the dependencies of $\lambda, \gamma, \delta$ into the expression for our list size proves \Cref{thm:fullfb}.

\section{Partial feedback setting (\Cref{thm:partfb})}\label{sec:partial_fb}

The protocol for the partial feedback channel is unrolled incrementally, with successive schemes layered on top of one another. We start in \Cref{subsec:partial_fb:rn} with a joint source-channel coding problem, with Alice having to communicate $\vbfx\in \alphX^N$, a noisy version of $\vbfy$, which Bob receives noiselessly. Alice leverages Bob's knowledge of $\vbfy$ to compress her vector $\vbfx$ into a smaller vector $\vbfz\in \alphX^{N\cdot H(\bfx|\bfy)}$. In \Cref{subsec:adv-noise}, we show how to equivalently express this setup as a channel coding problem with adversarial noise and add partial feedback. It then cleanly maps into the recursive formulation of the Weldon-type schemes discussed in \Cref{sec:full-fb}, with $\vbfx$ being a proxy for the $i^\text{th}$ stage $\stgi$, $\vbfy$ being Bob's received vector over the channel $\ystgi$, and $\vbfz$ being the follow-up transmission $\stgii$.

\subsection{Slepian-Wolf/Weldon schemes for random noise with minimal feedback}\label{subsec:partial_fb:rn}

\begin{problem}[Slepian-Wolf]
    \label[prb]{prb:random_errors}
    Let $\vbfy \in \alphX^N$ be a vector with $\bfy_i\sim P_\bfy$ drawn i.i.d. Alice observes a noisy version of $\vbfy$, denoted $\vbfx$, where $\bfx_i \sim P_{\bfx|\bfy}(\cdot|\bfy_i)$ i.i.d, for some conditional distribution $P_{\bfx|\bfy}$.
    Bob receives $\vbfy$. Alice wishes to transmit $\vbfz$, a compressed version of $\vbfx$, to Bob, such that Bob can reconstruct ${\vbfx}$ using $\vbfy$ and $\vbfz$ (see \Cref{fig:sw-setup}). The requirement is to design a \emph{computationally efficient} encoder, $\vbfz = f(\vbfx)$, and decoder $\hvbfx = g(\vbfy, \vbfz)$ such that 
    $\lim_{N\rightarrow\infty} \Pr\paren{\hvbfx\neq \vbfx} = 0$.
\end{problem}

\insertfig{fig:sw-setup}{
    The setup for \Cref{prb:random_errors}. The vectors $\vbfx, \vbfy$ share a joint distribution. Alice must prepare $\vbfz$ without observing $\vbfy$. Bob receives $\vbfz$ noiselessly and must recover $\vbfx$ from $\vbfy$ and $\vbfz$.
}{
    \includegraphics[trim=15mm 185mm 32mm 45mm, clip, width=0.99\linewidth]{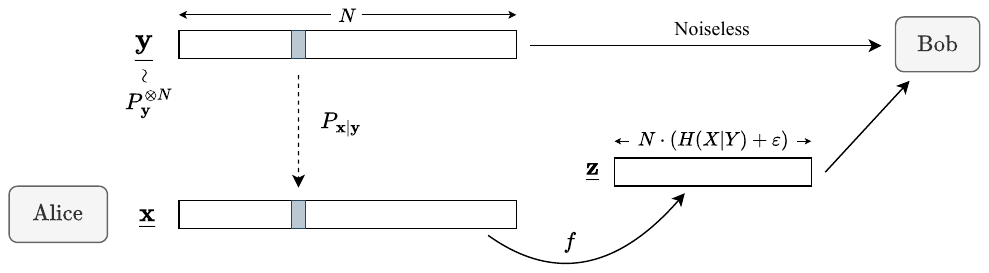}
}{
    Slepian-Wolf/Weldon-type schemes for random noise with minimal feedback (\Cref{subsec:partial_fb:rn})
}

\paragraph{Connection to Slepian--Wolf}
\Cref{prb:random_errors} is exactly the lossless source-coding-with-side-information setting: the decoder has noiseless access to $\vbfy$ and must recover $\vbfx$ from a compressed description $\vbfz=f(\vbfx)$. By the Slepian--Wolf theorem, the minimum achievable rate for $\vbfz$ is $H(\bfx|\bfy)$, and this corresponds to the corner point of the Slepian--Wolf region with $(R_X,R_Y)=(H(\bfx|\bfy),0)$ (equivalently, $R_Y$ is provided “for free” since $\vbfy$ is already known at the decoder).


{\bf (4.1.1) A computationally inefficient solution.} \phantomsection \label{para:sw}
We first develop a solution that is \emph{not} computationally efficient (C.E.). Alice uses a random hash function $h : \alphX^N \rightarrow \alphX^{N\cdot \paren{H(\bfx|\bfy)+\eps_h}}$ to encode and ``compress" $\vbfx$, and sends $\vbfz = h(\vbfz)$ to Bob noiselessly. To decode, Bob simply iterates through all vectors $\vu \in \alphX^N$, and checks whether it satisfies the hash function ($h(\vu)=\vbfz$), and that it is jointly typical with the previously received $\vbfy$. If he finds such a vector, he declares it to be $\hvbfx$, else he declares a decoding error. 

\begin{protocol}
\label{prot:sw_no-eff}
    Choose constant $\eps_h>0$, set $\eps_d = \eps_h/2$. A random hash function 
        $h : \alphX^N \rightarrow \alphX^{N\cdot \paren{H(\bfx|\bfy)+\eps_h}}$
    is generated (let $N' \coloneq N\cdot\paren{H(\bfx|\bfy)+\eps_h)}$ by assigning to each vector $\vu \in \cT^{(N)}_{\eps_d}({\bfx})$ a vector uniformly at random from $\alphX^{N'}$, and is shared with all parties. 
    Alice sends $\vbfz = h(\vbfx)$ to Bob. Bob iterates through all $\vu \in \Brackets{q}^N$ and checks for: 
    \begin{enumerate}
        \item Hash match: $h(\vu) = \vbfz$
        \item Joint typicality: $(\vbfy, \vu) \in \cT^{(N)}_{\eps_d}({\bfy, \bfx})$
    \end{enumerate}
    If there is a unique vector $\vu^*$ satisfying both conditions, Bob declares $\hvbfx = \vu^*$.
\end{protocol}

We show that this protocol meets all requirements of \Cref{prb:random_errors} in \Cref{subsec:app:sw-no-ce-proof}, with the exception of computational efficiency: Bob performs checks $\forall \; \vu \in \alphX^N$, taking exponential time. We reduce this to polynomial time, discussed ahead, and depicted in \Cref{prot:no-adv_ce}.

%

{\bf (4.1.2) Computational efficiency with concatenated codes.} \phantomsection \label{para:sw-concat}
In order to achieve C.E., we divide $\vbfx$ into smaller ``chunks" $\{\chni\}_{i=1}^K$ of length $\lchn = C_c\log(N)$, and apply $K$ hash functions $\{h_i\}_{i=1}^K$ (with appropriately scaled domain and range) separately on each chunk. Because the chunks are of length $C_c\log(N)$, Bob now only has to search through $\cO(2^{C_c\log(N)}) = \cO(N^{C_c})$ vectors  for each chunk, and since there are $K = \frac{N}{C_c\log(N)}$ chunks, the protocol runs in poly($N$) time.

It seems straightforward for Bob to now perform \Cref{prot:sw_no-eff} on each \emph{chunk} with encoders and decoders $(f_i,g_i)$, and append the reconstructed chunks $g_i(\ychni, \zchni)$ to get his estimate $\hvbfx$. However, from the analysis in \Cref{subsec:app:sw-no-ce-proof}, we can show that the probability that chunk $i$ is decoded incorrectly is
\begin{align}
    P_{e}^{(i)} &\coloneqq \Pr\parens*{g_i\parens*{\ychni, \zchni} \neq \chni } \\
    &\leq 2^{-C_c\log(N)\eps_h/3} = N^{-C_c\eps_h/3}, \label{eq:pei}
\end{align}
We use a systematic Reed-Solomon outer code over alphabet size $Q=q^{\lchn}=q^{C_c\log(N)}=N^{C_c}$, (systematic implying the first $K$ chunks of the encoded vector are equal to the first $K$ chunks of $\vbfx$). 

\begin{protocol}[Random errors, Computationally Efficient]
    \label{prot:no-adv_ce}
    \
    \greyhrule
    \noindent\textbf{Shared Setup}
    Choose values $C_c, \eps_h, \eps_d, \eps_H, K'$ as detailed in \Cref{subsec:app:no-adv-ce-proof}. The joint distribution $P_{\bfx,\bfy}$ is known to all parties. Let $\lchn\coloneq C_c\log(N)$ and $K\coloneq\nicefrac{N}{\lchn}$. A set of hash functions $\cH = \left\{ h_i: \Brackets{q}^\lchn \rightarrow \alphX^{\lchn\cdot\paren{H(\bfx|\bfy)+\eps_h}} \right\}_{i=1}^{K}$ is shared with all parties.
    Fix a $(K',K)$ Systematic Reed Solomon code over $\bF_Q$ ($Q=N^{C_c}$) with encoder and decoder $\encRS, \decRS$ respectively.
    \greyhrule
    \noindent\textbf{Encoding:}
    Alice partitions $\vbfx$ into $K$ chunks, $\{\chni\}_{i=1}^K$, each of length $\lchn$. She applies the encoder chunk-wise:
    \begin{equation*}
    \begin{split}
        \parens*{\chnone', \chntwo', ... , \chnkd'} \\ =\encRS\parens*{\chnone,\chntwo, ... ,\chnk}.
    \end{split}
    \end{equation*}
    Alice compresses to $\vbfz$ using the hash functions (see \Cref{fig:no-adv_ce}):
    \begin{equation*}
    \begin{split}
        &\zchni =
        \begin{cases}
            h_i\parens*{\chni'}, &i\in[K]\\
            \chni', &i\in[K']\setminus[K],
        \end{cases} \\
        &\vbfz = \parens*{\zchni}_{i=1}^{K'}.
    \end{split}
    \end{equation*}
    \greyhrule
    \noindent\textbf{Decoding:} For each chunk $i\in[K]$, Bob iterates through all $\vu \in \Brackets{q}^{\lchn}$ and checks:
    \begin{enumerate}
        \item Hash match: $h_i(\vu) = \zchni$
        \item Joint Typicality: $\parens*{\ychni, \vu} \in \cT_{\bfy,\bfx}^{(\lchn)}(\eps_d)$, for $i\in\{1,2, ... ,K\}$
    \end{enumerate}
    \noindent If $\exists$ a unique $\vu\in\alphX^{\lchn} \;\forall i\in[K]$ satisfying these conditions, Bob assigns that value to $\hvbfx_{(i)}$, and declares $\hvbfx = \decRS\parens*{\hvbfx_{(1)}, \hvbfx_{(2)}, ... , \hvbfx_{(K)}, \vbfz_{(K+1)}, ... , \vbfz_{(K')}}$.
    %
    \greyhrule
    \noindent\textbf{Performance:}
    The protocol completes with $P_e = \Pr(\hvbfx \neq \vbfx) \leq \exp\parens*{-\dfrac{1}{3}\cdot\frac{N^{1-C_c\eps_H}}{C_c\log N\paren{1-4N^{-C_c\eps_H}}}}$
\end{protocol}

Proof of performance and details of the protocol are in \Cref{subsec:app:no-adv-ce-proof} and \Cref{tab:constants}). 

\insertfig{fig:no-adv_ce}{
    \Cref{prot:no-adv_ce}: A systematic R.S code acts on $\vbfx$ by treating each chunk of length $C_c\log(N)$ as an element from an alphabet of size $N^{C_c}$. The systematic chunks are compressed by a hash function and the redundancy chunks are sent as is. Breaking up the vector $\vbfx$ into chunks of length $\Theta(\log(N))$ allows us to achieve computational efficiency.
}{
    \includegraphics[trim=33mm 185mm 20mm 31mm, clip, width=0.99\linewidth]{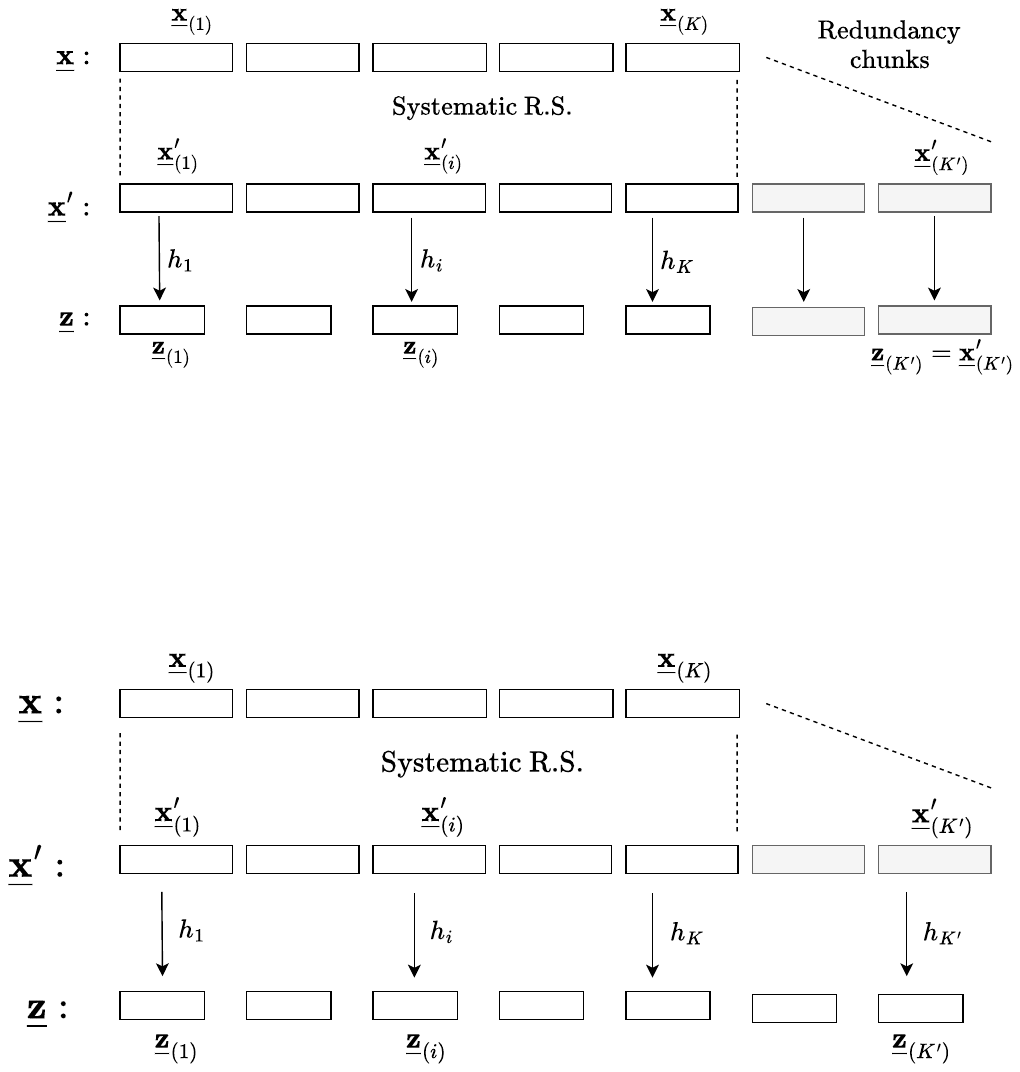}
}{
    Computational efficiency with concatenated codes (\hyperref[para:sw-concat]{Section 4.1.2})
}
\subsection{Adversarial noise}
\label{subsec:adv-noise}

We now re-cast the source coding problem to fit into our feedback channel setting -- see \Cref{fig:noise-to-adv-flip}. Instead of $\vbfx$ being generated from $\vbfy$, the action of a noisy channel generates $\vbfy$ from $\vbfx$. If this channel were a DMC, the two settings are immediately equivalent. We upgrade the channel to being adversarially controlled by James, and similarly upgrade our protocols to be able to handle adversarial action. 

\begin{remark}
\label{rem:weldon-mapping}
    The channel coding setup in \Cref{fig:noise-to-adv-flip} is then readily mapped into the Weldon-type framework: $\vbfx$ is the proxy for the $i^\text{th}$ stage $\stgi$, $\vbfy$ is what Bob receives, $\vbfy^{(i)}$, and $\vbfz$ is the transmission for the \emph{next} stage, $\stgii$.
    Note that we end the entire Weldon-type protocol with a termination stage (described in \hyperref[para:term_scheme]{Section 3.1.3}), from which Bob decodes the last stage reliably. Thus reconstruction of $\vbfx$ given noiseless reception of $\vbfz$ implies that we can apply this setup recursively on every stage of the Weldon scheme, starting backwards from the last stage, and successfully recover the original message $\vbfm$.
\end{remark}

 We retain the notation $\vbfx, \vbfy, \vbfz$ instead of $\stgi, \ystgi, \stgii$ for brevity. We first begin with James being given ``semi-adversarial" powers, and scale up from there.

\insertfig{fig:noise-to-adv-flip}{
    The mapping between Slepian-Wolf source coding, and channel coding. We ``replace" the distortion in the form of the joint distribution by an adversary James. This makes for a clean mapping to the channel setting, and a further mapping to the recursive Weldon-type scheme apparent. If we can demonstrate success of this toy example by receiving $\vbfz$ noiselessly, the same protocol can be applied iteratively.
}{
    \includegraphics[trim=30mm 147mm 54mm 40mm, clip, width=0.99\linewidth]{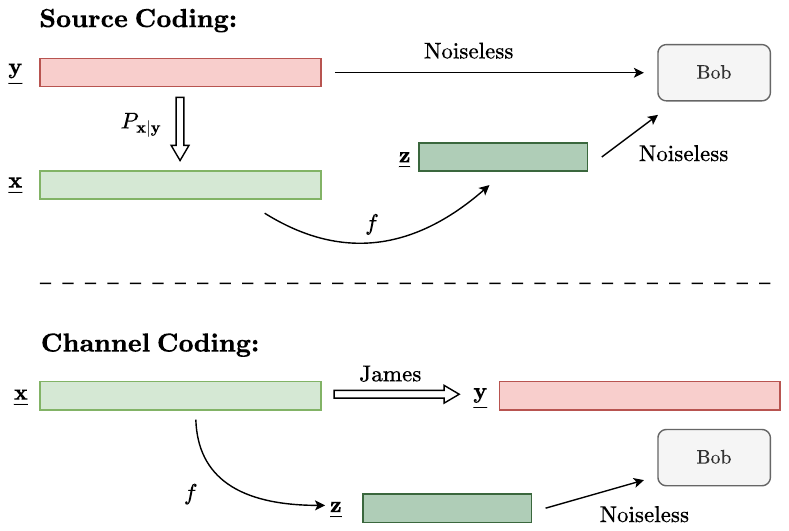}
}{
    Adversarial noise (\Cref{subsec:adv-noise})
}

\begin{problem}[Semi-Adversarial Noise]
    \label[prb]{prb:semi-adv}
    Alice wishes to communicate $\vbfx \in \alphX^N$ to Bob. James chooses $\vbfs \in \alphS^N$, and Bob observes the channel output $\vbfy = \vbfx \oplus \vbfs$. In the semi-adversarial setting, the channel is a $q$-SC with crossover probability $p$, but the value of $p$ is chosen adversarially by James (potentially as a function of $\vbfx$) and is unknown to all other parties, i.e. $\Pr(\bfs_i=j)=(1-p)$ if $j=0$ and $\nicefrac{p}{(1-q)}$ if $j\neq 0$.
    Alice receives feedback $\vbfF = \fbfunc(\vbfy)$ from Bob, and prepares a ``clean up" transmission $\vbfz = f(\vbfx, \vbfF)$ which Bob receives \emph{noiselessly}. Bob decodes his estimated vector $\hvbfx = g(\vbfy, \vbfz)$. The goal is to construct a \emph{computationally efficient} encoder and decoder pair $(f, g)$ such that
        $\lim_{N\rightarrow\infty}\Pr(\hvbfx \neq \vbfx) = 0$.
\end{problem}

    If Alice knows $p$ this  reduces to \Cref{prb:random_errors} solved C.E. by \Cref{prot:no-adv_ce}.
We can give Alice knowledge of $p$ by granting \textbf{partial feedback} to Bob. Bob simply sends back the first $C_e\log(N)$ symbols of $\vbfy$, i.e. $\vbfF = \vbfy_{1:C_e\log(N)}$, which Alice compares with $\vbfx$. Since the noise is i.i.d, Alice obtains a reliable estimate via a plug-in estimator:
$    \tilde{p} = \dfrac{1}{C_e\log(N)}\sum_{j=1}^{C_e\log(N)} \mathds{1} \left\{\bfx_j\neq\bfF_j\right\}.$
By the Chernoff bound, it can be shown that
   $ \Pr\parens*{|\tilde{p}-p|>\eps_e} \in \cO\paren{N^{-C_e\eps_e^2}}.$
Armed with a good estimate for $p$, Alice then carries out \Cref{prot:no-adv_ce}.
\begin{remark}
    In the semi-adversarial setting, Alice does not know the value of $p$ beforehand, and so the range of the hash function cannot be decided prior to the start of the protocol. To get around this, \Cref{prot:no-adv_ce} is further modified by sharing the hash set $\cH = \{h_i:\Brackets{q}^\lchn\rightarrow \Brackets{q}^\lchn\}_{i=1}^K$ {\it a priori}. After getting an estimate $\tilde{p}$, Alice truncates the range of all $h_i(\cdot)$ to the required length $\ell_c\parens*{H_q(\tilde{p})+\eps_h}$.
\end{remark}

\begin{problem}[Fully Adversarial Noise]
    \label[prb]{prb:full-adv}
    Alice sends $\vbfx\in\alphX^N$, Bob receives $\vbfy = \vbfx \oplus \vbfs \in \alphX^N$, where $\vbfs$ is generated adversarially by James. Alice receives feedback $\vbfF = \fbfunc(\vbfx)$ from Bob, and prepares her next transmission $\vbfz = f(\vbfx, \vbfF)$, which Bob receives \emph{noiselessly}. Bob needs to generate a reliable estimate $\hat{\vbfx} = \hat{\vbfx}(\vbfy, \vbfz)$.
\end{problem}


{\bf (4.2.1) Estimating noise levels.}\phantomsection\label{para:est-nl} 
Now we successively layer tools onto the previous protocol to reduce this problem back to \Cref{prb:semi-adv}.
Alice first attempts to estimate the type-class of $\vbfs$. Instead of simply sending the first $C_e\log(N)$ symbols of $\vbfx$, Bob now randomly samples a subset of indices $T\subset [N]$ of size $|T| = C_e\log(N)$. Bob sends his feedback $\vbfF$ back as $\Fa = T$, the set of indices, and $\Fb = \vbfy_T$, i.e. the symbol values at the indices.

Alice now knows $C_e\log(N)$ symbols (and their positions) as received by Bob. She compares those symbols in $\vbfy$ with  corresponding symbols in $\vbfx$ to get an estimate of the noise levels:

\begin{equation}
\label{eq:est-noise}
    \tilde{p} = \dfrac{1}{|T|} \sum_{j=1}^{|T|} \one\left\{ \vbfF_{(b),j} \neq \bfx_{T_j} \right\} 
\end{equation}.

From \Cref{thm:learning-distr-linfty} it follows directly that $    \Pr\parens*{|\tilde{p}-p|\geq\eps_e} \leq 2N^{-C_e\eps_e^2/2}$.

\begin{remark}
    Sending the set $T$ would take Bob an additional $\log(N)\cdot C_e\log(N)=C_e\log^2(N)$ symbols.
\end{remark}

In summary, Alice now has an accurate estimate $\tilde{p}$ of the noise level James has used.
This solution, however, is not complete. For the chunking and hashing procedure to work as it did in \Cref{prot:no-adv_ce}, each chunk of $\vbfx$ that Alice hashes must have an i.i.d distribution of noise, however in the current setting, $\vbfs$ is chosen adversarially by James. The tools introduced ahead compensate for this mismatch.

{\bf (4.2.2) Permutations.}\phantomsection\label{para:perm}
In order for each chunk that Alice makes to have roughly $p$ fraction of errors in them, Alice selects chunks by \emph{permuting} $\vbfx$ randomly, as depicted in \Cref{fig:perms}.
Of course, James must not know the permutation while he's committing errors, otherwise he could strategically plant his errors such that, after applying the permutation, the errors are still distributed in a skewed manner across chunks.

Let $\cP = \{\perm^1, \perm^2, ..., \perm^{|\cP|}\}$ be a set of $|\cP|=N^{C_p}$ permutations on $N$, with $\perm^i:[N]\rightarrow[N]$ and
$
    \perm^i(\vbfx) \coloneq \parens*{\bfx_{\perm^i(1)}, \bfx_{\perm^i(2)}, ... , \bfx_{\perm^i(N)}}.
$
Bob sends Alice $\Fc \sim \unif(\alphX^{C_p\log(N)})$ as feedback to select which permutation Alice will use to prepare her transmission $\vbfz$. Let $I\parens{\Fc}\in\bracks*{N^{C_p}}$ be the integer index that $\Fc$ corresponds to. Alice generates $\permed = \perm^{I\parens*{\Fc}}(\vbfx)$, the ``shuffled" vector.

Bob's feedback $\vbfF$ is sent \emph{after} James commits $\vbfs$ (after Bob receives $\vbfy=\vbfx \ \oplus\ \vbfs$). This ensures James cannot plan his error patterns with a particular permutation in mind.
\Cref{lem:pgood} shows that a large fraction of chunks are indeed \emph{quasi-uniform}, i.e. they have roughly $p$ fraction of symbol flips w.h.p., however they may not be distributed within the chunk in an i.i.d manner (as was required in \Cref{prot:no-adv_ce}).

\insertfig{fig:perms}{
    Alice applies a permutation to create her ``shuffled" vector $\permed$
}{
    \includegraphics[trim=45mm 186mm 65mm 56mm, clip, width=0.99\linewidth]{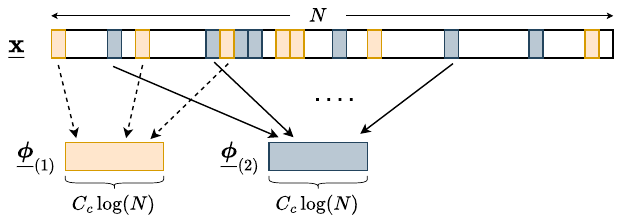}
}{
    Permutations (\hyperref[para:perm]{Section 4.2.2})
}

{\bf (4.2.3) Random hashes.}\phantomsection\label{para:hash} 
To address the problem of quasi-uniformity within each chunk, we equip the hash from \Cref{prot:no-adv_ce} with a short, per-chunk random seed. Bob draws a seed $\rchni \sim \unif\parens*{\alphX^{\sqlch}}$ and sends it as feedback. Alice computes $\zchni = h\parens{\permed_{(i)}, \rchni}$ using a random hash function $h: \alphX^\lchn\times\alphX^{\sqlch} \rightarrow \alphX^{\lchn (H_q(\tilde{p})+\eps_h)}$.
%
%
Intuitively, the extra $\sqrt{\lchn}$ symbols per chunk select independent hash tables across seeds, so only a vanishing fraction of seeds are ``bad", i.e. result in a hash collison. Picking the seed uniformly at random, then, makes the per-chunk error probability exponentially small in $\sqlch$.

For a given instantiation of a chunk (say chunk 1) and it's corresponding attack vector, $\parens{\vpermed_{(1)}, \vs_{(1)}}$, define a ``bad" seed $\tvr$ as one that causes a hash collision, i.e. there exists $ \tilde{\vpermed}_{(1)} \in \alphX^{\lchn}$ and a distinct $\tilde{\vpermed}_{(1)} \neq \vpermed_{(1)}$ such that $\tilde{\vpermed}_{(1)}$ is in the Hamming ball $\cB\parens{\vy_{(1)}, \lchn p(1+\eps_d)}$, and the two hashes $h\parens{\tilde{\vpermed}_{(1)}, \tvr}$ and $h\parens{\vpermed_{(1)}, \tvr}$ are equal.
Now, let $\bfB\parens{\vpermed_{(1)}, \vs_{(1)}}$ represent the number of bad seeds $\tvr$ from a total of $q^{\sqlch}$ possible seeds. \Cref{lem:few-bad-seeds} shows that $\bfB\parens{\vpermed_{(1)}, \vs_{(1)}} \leq q^{\sqlch/2}$ with probability tending to one. Picking a seed uniformly at random hence ensures a good seed w.p. $1-q^{-\sqlch/2}$. This result holds for any chunk $i$; i.e. $P_e^{(i)} = q^{-\sqlch/2}$. These errors are independent across chunks since $\vbfr_{(i)}$ are generated i.i.d.
Since each chunk needs $\sqrt{\lchn} = \sqrt{C_c\log N}$ symbols as input to the hash, and there are $K = N/(C_c\log N)$ chunks, Alice requires $K\cdot\sqrt{\lchn} = N/(\sqrt{C_c\log N})$ random symbols from Bob as feedback in order to choose the seeds for her hashing procedure.
With similar motivations as in \Cref{prot:no-adv_ce}, we use a systematic Reed Solomon outer code to ``clean up" the bad chunks caused both by the permutation bank and the hash function.
Compiling these discussions, we outline the solution to \Cref{prb:full-adv} (the proof of performance of this protocol is detailed in \Cref{subsec:app:adv-ce-proof}). See \Cref{tab:constants} for a guide to constants.

\subsection{\Cref{prot:adv-ce}}

\begin{protocol}[Adversarial Errors, Computationally Efficient]
    \label{prot:adv-ce}
    \
    \greyhrule
    \noindent\textbf{Shared Setup}
    Pick values $C_e, \eps_e, C_c, C_p, \eps_N, \eps_h, \eps_d, K'$ as detailed in \Cref{subsubsec:app:adv-ce-params}. Define $\lchn\coloneqq C_c\log(N),\; K\coloneqq\nicefrac{N}{\lchn}$. Hash table $h:\alphX^\lchn\times\alphX^{\sqrt{\lchn}} \rightarrow \alphX^\lchn$ and permutation list $\cP=\{\perm^i\}_{i=1}^{(N^{C_p})}$ are shared with all parties. Fix a $(K',K)$ systematic RS code over $\bF_Q$, where $Q=N^{C_c}$.
    \greyhrule
    \noindent \textbf{Channel:}
    Alice sends $\vbfx$, James chooses $\vbfs$ and Bob receives $\vbfy = \vbfx \oplus \vbfs$.
    \greyhrule
    \noindent \textbf{Feedback:}
    Bob uniformly samples (i.i.d) a set of indices $T\subset[N], \: |T|=C_e\log(N)$, and prepares $\Fa = T$, $\Fb = \vbfy_T$:
    \begin{equation*}
    \begin{split}
        &\Fc \sim \unif\parens*{\alphX^{C_p\log(N)}}, \\
        &\Fd \sim \unif\parens*{\alphX^{{N}/{\sqrt{C_c\log N}}}}.
    \end{split}
    \end{equation*}
    Bob sends $\vbfF = \parens*{\Fa, \Fb, \Fc, \Fd}$ noiselessly to Alice.
    \greyhrule
    \noindent \textbf{Encoding:} Alice estimates $\tilde{p}$ using \Cref{eq:est-noise}. She chooses a permutation $\rperm=\perm^{I\parens*{\Fc}}$ and generates $\permed = \rperm(\vbfx)$. Divide $\permed$ into $K$ chunks $\parens*{\permed_{(i)}}_{i=1}^K$, each of length $\lchn$.
    Applying the RS Code yeilds:
    \begin{equation*}
    \begin{split}
        &\parens*{\permed_{(1)}', \permed_{(2)}', ... , \permed_{(K')}'} \\
        &= \encRS\parens*{\permed_{(1)},\permed_{(2)}, ... ,\permed_{(K)}}.
    \end{split}
    \end{equation*}
    Alice obtains random seeds by diving $\Fd$ into equal partitions of length $\sqrt{\lchn}$, i.e. 
    \begin{equation*}
        \vbfr_{(i)} = \parens*{\Fd}_{(i-1)\sqrt{\lchn}+1:i\sqrt{\lchn}} \quad ,\quad i\in[K].
    \end{equation*}
    Alice truncates the hash table to have range $\lchn (H_q(\tilde{p})+\eps_h)$, and compresses:
    \begin{equation*}
    \begin{split}
        &\zchni =
        \begin{cases}
            h\parens*{\permed_{(i)}, \vbfr_{(i)}}, &i\in[K] \\
            \permed_{(i)}, &i\in[K']\setminus[K],
        \end{cases} \\
        &\vbfz = \parens*{\zchni}_{i=1}^{K'}.
    \end{split}
    \end{equation*}
    \greyhrule
    \noindent\textbf{Decoding:} Bob receives $\vbfz$ noiselessly and obtains the value of $\tilde{p}$. Bob generates the permuted $\ypermed = \rperm(\vbfy)$ and chunks it. For each chunk $i\in[K]$, Bob checks, for each $\vu\in\alphX^{\lchn}$
    \begin{enumerate}
        \item Hash match: $h(\vu)=\zchni$
        \item Typicality: $\dh\parens*{\vu, \ypermed_{(i)}} \in [\lchn(\tilde{p}-\eps_d), \lchn(\tilde{p}+\eps_d)]$.\label{item:typicality-test}
    \end{enumerate}
    If a unique vector $\vu^*$ exists he assigns it to the estimate $\hat{\permed'}_{(i)}$ for $i\in[K]$, and declares $\hat{\permed} = \decRS\parens*{\hat{\permed'}_{(1)}, \dots ,\; \hat{\permed'}_{(K)}, \;\vbfz_{(K+1)}, ... , \;\vbfz_{(K')}}$,
    and finally $\hvbfx = \rperm^{-1}\parens*{\hat{\permed}}$
    \greyhrule
    \noindent\textbf{Performance:} The protocol completes with $P_e\leq 2\parens*{N^{-\eps_p} + N^{-2\eps_e^2C_e}}$.
\end{protocol}

\section{Partial feedback: Putting it all together}\label{sec:compile}

The progression of problems and protocols is depicted in \Cref{fig:sol-overview}. \Cref{prb:full-adv} maps directly into the Weldon-type structure of the true protocol, as discussed in \Cref{rem:weldon-mapping}. The termination scheme is identical to the one detailed in \hyperref[para:term_scheme]{Section 3.1.3}.
%
%
\insertfig{fig:sol-overview}{
    The overview of the solution for the partial feedback regime. This figure depicts the various tools and conditions layered onto the Slepian-Wolf problem to gradually upgrade it until it can handle adversarial error patterns.
}{
    \includegraphics[width=0.9\linewidth, trim = 45mm 85mm 50mm 25mm , clip]{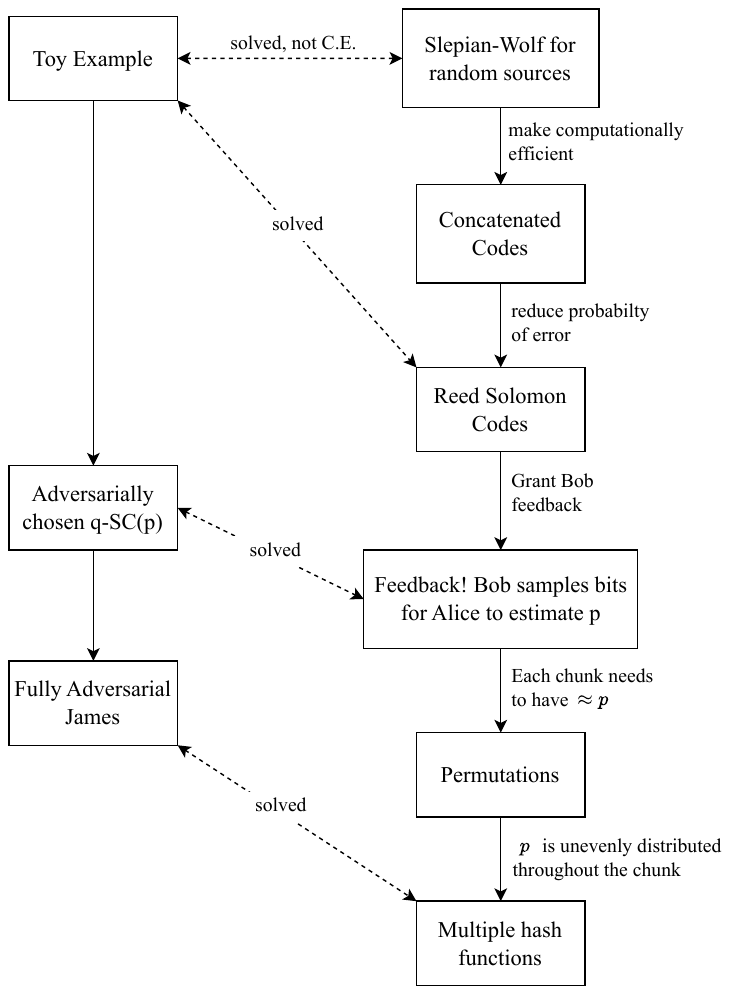}
}{
    Partial feedback: Putting it all together (\Cref{sec:compile})
}
\subsection{Synchronization}
\label{subsec:partialfb-sync}

Bob needs to synchronize two aspects of the protocol. The first, discussed in \Cref{subsec:fullfb-sync}, is that \underline{Bob is not aware of James' error fraction} in each stage, hence cannot isolate the stages in order to decode them. Bob guesses, by brute force, the error vector $\hat{\vp} = (\hp_1, \hp_2,..., \hp_{\nstages})$, with $\hp_i \in \{0, \delta, 2\delta, \dots 1 - \delta, 1\}$. We are guaranteed that one of the at most $(1/\delta)^\lambda$ guesses is the correct error fraction vector. The implications of this discretization on the choice of constants for the scheme is outlined in \Cref{subsec:app:adv-ce-fullscheme-constants}. Note that the dependency of the overall list size $|\cL|$ on $\eps$ is dominated by this guessing procedure in the Weldon-type phase, as opposed to the list size generated by the termination scheme.

The second synchronization is required since, in the full feedback setting, Bob was immediately feeding back every received symbol. Then, in \Cref{prb:semi-adv,prb:full-adv} Bob knew exactly when to send feedback to Alice (after $\vbfx$ is transmitted). In our Weldon scheme, however, Bob is not aware of the demarcations for each stage, neither is he allowed to send the symbol as feedback immediately (only partial, periodic feedback is allowed); hence \underline{he does not know when to send feedback}. To work around this, we have Bob send his feedback vector $\vbfF$ after every \textbf{block} of length $n\kappa$ that he receives from Alice, regardless of the stage length. Alice pads the end of her stages with zeros to ensure that its length is a multiple of $n\kappa$, see \Cref{fig:sync}. From initial blocks, Alice only keeps the $C_e\log(n\kappa)$ symbols she needs to get her noise estimate $\tilde{p}$. From the block on which her stage also ends, she selects all feedback vectors and uses them to select the permutations and hash tables for the current stage. The intended length of the $i$th stage is $\ell_i$, however the resultant length is $\hell_i=\ceil*{\frac{\ell_i}{n\kappa}}\cdot n\kappa$.

\insertfig{fig:sync}{
    Alice and Bob synchronize stage endings. Alice waits to end her stage on a multiple of $n\kappa$ (padding with zeros, as shown in the figure), so she can use the feedback vectors $\Fc$ and $\Fd$ generated in the most recent $n\kappa$ block. All the feedback that Alice uses is shown selected in the dashed grey box. Alice discarding so much feedback may seem wasteful, but recall that each of these vectors is $o(n)$ length.
}{
    \includegraphics[trim = 45mm 155mm 67mm 68mm, clip, width=0.99\linewidth]{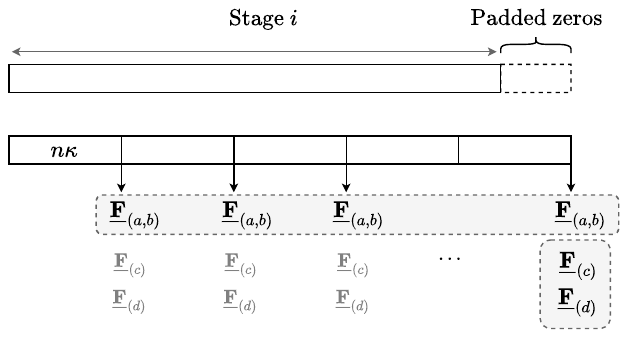}
}{
    Amount of feedback (\hyperref[para:params]{Section 5.1.1})
}

{\bf (5.1.1) Choice of parameters.}\phantomsection\label{para:params} 
We choose the same asymptotic dependencies for $\delta$ and $\tilde{\lambda}$ (where $\tilde{\lambda}$ is an upper bound on $\lambda$) on $\eps$ and $\gamma$ as in the full feedback scheme. The fact that every stage length is being rounded up to a multiple of $n\kappa$ mandates a careful choice of $\kappa$ to ensure we terminate within $n$ forward transmissions. We show in \Cref{subsec:partfb-recursion}, by modifying the analysis done for the full feedback setting in \Cref{subsec:app:clean-descent}, that choosing $\kappa=\Theta\parens*{ \eps^{\tilde{\lambda}+\nicefrac{3}{2}} }$ ensures this. These alterations allow us to fit \Cref{prot:adv-ce} cleanly into the recursive Weldon-type formulation. The protocol performance metrics, such as the probability of error, list size, and storage costs are detailed in \Cref{subsec:app:adv-ce-fullscheme-constants}.

\textbf{Amount of Feedback:} Note $\vbfF_{(a)}, \vbfF_{(b)}$ are meant to help Alice estimate James' noise levels in that block, hence they are of size $C_e\log^2(n\kappa), \;C_e\log(n\kappa)$ respectively. Alice can then combine them to estimate $\hp_i$ for the stage.
Since we know, for every stage, $N = \ell_i \leq nR < n$, it suffices to send $C_p\log(n)$ random symbols for $\vbfF_{(c)}$ and $\frac{n}{\sqrt{C_c\log(n)}}$ random symbols for $\vbfF_{(d)}$. This allows Alice to simply select $\vbfF_{(c)}, \vbfF_{(d)}$ from the very last block -- the one that she sets her stage to end on -- and use those to select the random permutation and random hashes for the entire stage, discarding the rest (\Cref{fig:sync}). To complete the proof of \Cref{thm:partfb}, we note that over the entire scheme, our total number of feedback symbols are
\begin{align*}
   \frac{1}{\kappa} \cdot \Big(C_e\log(n\kappa) &+ C_e\log^2(n\kappa) + C_p\log(n) \\
   &\hspace{50pt}+ \frac{n}{\sqrt{C_c\log(n)}}\Big) \in o(n).
\end{align*}

\section{Conclusion}
We have provided the first computationally efficient algorithm for the adversarial list decoding problem with feedback that decodes up to information-theoretic limits, in both the full feedback setting and the vanishing feedback setting. Some possible future directions of research include investigating whether these results can be extended from the $q$-ary bit flip model we considered to the more general class of \emph{arbitrarily varying channels} (AVCs), establishing whether there exists an efficient \emph{deterministic} coding scheme that approaches information-theoretic limits in the vanishing feedback setting, and investigating how small the vanishing fraction of feedback can be made while still achieving the optimal rate of $1-H_q(\varrho)$.

\section*{Acknowledgments}

The authors would like to acknowledge helpful exchanges with (in alphabetical order) Nicolas Resch, Atri Rudra and Mary Wootters.

\section*{AI Use Statement}
This work was done without the use of AI—all results were derived and written by the authors.


\appendix

\iffiguresatend
  \section{Figures}
  \myfigurelist
\fi
\subsection{Recursive bounds on Weldon-type scheme parameters in adversarial noise settings}
\label{sec:app:descent}

\begin{figure}[h]
    \centering
    \includegraphics[trim=36mm 175mm 45mm 30mm, clip, width=0.99\linewidth]{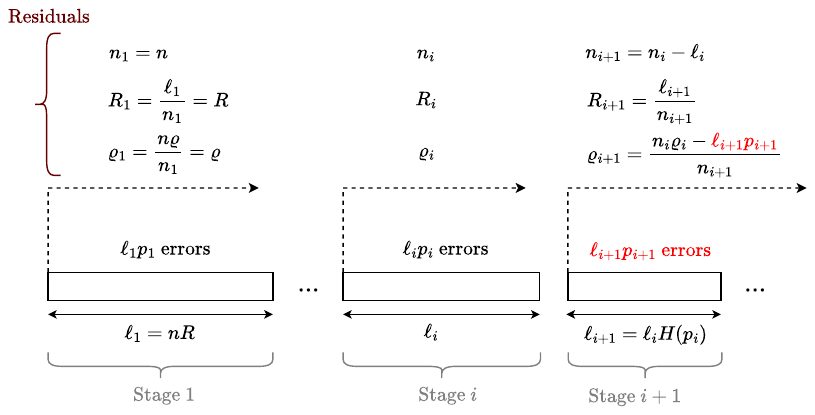}
    \caption{Notation employed in the recursive Weldon-scheme for adversaries}
    \label{fig:weldon-fullfb-notation}
\end{figure}

Recall the definitions of various recursive quantities. The length of the $i$th stage is $\ell_i\coloneq |\vx^{(i)}|$. The remaining transmissions Alice has remaining at the onset of stage $i$ is denoted $n_i$, and the fraction of $n_i$ that the current stage occupies is the ``residual rate", $R_i\coloneqq \ell_i/n_i$. James flips a \emph{fraction} $p_i$ transmissions of stage $i$ (hence the actual number of errors used in that stage are $\ell_ip_i$). The ``residual budget" $\varrho_i$ is defined as the amount of flips James has available to him at the onset of stage $i$, as a fraction of the remaining blocklength $n_i$. See \Cref{fig:weldon-fullfb-notation} for a visual depiction.

\noindent Fix the rate as $ R = 1 - H_q(\varrho) - \eps $. James can flip a total of $n\varrho$ transmissions across the entire protocol. Consider a message $ m \in \Brackets{q}^{nR} $ and let the initial stage be $ \vx_1 = m$ -- hence $\ell_1\coloneqq |\vx_1| = nR$ and $n_1$, the number of channel uses left at the onset of stage 1, is equal to $n$. 
The pair of residual rate and residual budget are hence initialized as 
\begin{align}
&&
    R_1 &\coloneq \frac{\ell_1}{n_1} = R, &
    \varrho_1 &\coloneq \frac{n\varrho}{n_1} = \varrho . & 
& \notag 
\end{align}

\noindent For each stage $ i\ge1 $, repeat the following two steps until we hit the termination condition (see \hyperref[para:term_scheme]{Section 3.1.3}): 
\begin{enumerate}
    \item Alice transmits the length-$ \ell_i $ sequence $ \vx_i $ in which a fraction of $ p_i $ symbols are corrupted by James and Bob receives the resulting $ \vy_i $. 
    Then the residual blocklength is updated as $ n_{i+1} = n_i - \ell_i $. 

    \item Alice constructs $\vx_{i+1}$ as a sequence of length $\ell_{i+1} = f(\ell_i,p_i)$. The residual rate $R_{i+1}$ and residual budget $\varrho_{i+1}$ are updated as:
    \begin{align} 
        R_{i+1} &\coloneq \frac{\ell_{i+1}}{n_{i+1}} = \frac{f(\ell_i,p_i)}{n_i-\ell_i} \\ 
        \varrho_{i+1} &= \frac{\varrho_i - R_i p_i}{1 - R_i}, 
 \label{eqn:R_rho_update}
    \end{align}
\end{enumerate}
where \Cref{eqn:R_rho_update} follows from the definition of $R_i$ and the fact that James' total budget must be conserved,
\begin{align*}
    n_i\varrho_i &= \ell_ip_i + n_{i+1}\varrho_{i+1}\\
    \implies \varrho_i &= R_ip_i + (1-R_i)\varrho_{i+1}.
\end{align*}
Convexity of the mutual information functional $ x \mapsto 1 - H_q(x) $ implies: 
\begin{align}
    (1 - R_i) (1 - H_q(\varrho_{i+1})) &+ R_i (1 - H_q(p_i)) \\
    &\hspace{40pt}\ge 1 - H_q(\varrho_i) . \notag 
\end{align}
Defining the gap-to-capacity in stage $i$ as $ \eps_i \coloneqq 1 - H_q(\varrho_i) - R_i $ (with $\eps_1 \coloneqq \eps$), and substituting for $H_q(\varrho_i), H_q(\varrho_{i+1})$, we are led to a recursive inequality: 
\begin{equation}
    \label{eq:eps-recur}
    \eps_{i+1} \geq \frac{\eps_i}{1-R_i} - \parens*{R_{i+1} - \frac{R_iH(p_i)}{1-R_i}}
\end{equation}
The value of $R_{i+1}$, from \Cref{eqn:R_rho_update}, depends on how one chooses to encode the next chunk from the previous one. We analyze the behaviour of the recursion for the values of $R_{i+1}$ emergent from our schemes.

\subsection{Weldon-type recursions with no slack factors}
\label{subsec:app:clean-descent}

James corrupts a fraction $p_i$ of transmissions from the $i$th chunk $\vbfx_i$ of length $\ell_i$. To begin, we analyze an idealized ``clean recursion", where Alice constructs the next chunk of length $\ell_{i+1}=\ell_i H_q(p_i)$ - from \Cref{eq:clean-recurs-ell-update} - to describe the error pattern. We have
\begin{equation*}
    R_{i+1} = \frac{\ell_{i+1}}{n_{i+1}} = \frac{\ell_i H_q(p_i)}{n_i - \ell_i} = \frac{R_iH_q(p_i)}{1-R_i}.
\end{equation*}
Substituting the value of $R_{i+1}$ into \Cref{eq:eps-recur}:
\begin{equation}
    \label{eq:eps_i_bound}
    \eps_{i+1}\geq \frac{\eps_i}{1-R_i}.
\end{equation}
Unrolling this recursion gives
\begin{align}
   \label{eq:eps_bound_final} \eps_{i+1} &\ge \eps \prod_{j = 1}^i \frac{1}{1 - R_j}.
\end{align}
Note that this recursive formula also implies that the sum of the stage lengths $\ell_i$ before termination is strictly less than the total blocklength $n$ (we will separately discuss the stage length of the termination scheme shortly). To see this, recall the definition of the residual blocklength $n_i = n_{i-1} - \ell_{i-1}$ for $i \geq 2$ (with $n_1 = n$), where $\ell_i$ is the length of stage $i$. By definition, $n_i$ is the remaining number of transmissions until the $n$-th transmission entering stage $i$. Thus, to show that the sum of the stage lengths does not exceed $n$, it suffices to show that $n_i > 0$ for each $i$ (since this would imply that $\ell_i < n_i$ for all $i\geq 1$).
To do so, consider an arbitrary stage $i$ with blocklength $n_i > 0 $. In this stage, Alice has residual rate $R_i$, and sends her unencoded message of length $n_iR_i$ across the channel. Accordingly, the residual blocklength in stage $i+1$ is then $n_{i+1} = n_i(1-R_i)$, giving us a recursive relation for $n_i$ in terms of $n_{i-1}$ and $R_{i-1}$. Since $R_i \leq 1 - \eps_i$ and by the bound $\eps_i \geq \eps$ due to \eqref{eq:eps_i_bound}, we can then lower bound $n_{i+1}$ by $n_{i+1} \geq \eps > 0$. Hence, if $n_i > 0$, then the same holds for $n_{i+1}$. Using the fact that $n_1 = n > 0$ trivially, the above holds for all $i \geq 1$ by an induction argument, implying our claim on the sum of the stage lengths.

The recursive formulas for $\eps_i$ in \eqref{eq:eps_i_bound} and \eqref{eq:eps_bound_final} also put us in a position to bound the number of stages until termination. Recall that the termination condition is for the residual rate to lie below the Zyablov bound:
\begin{equation}
    \label{eq:zyablov_def} R_{\rm Z}(\varrho) =  \underset{0 < r < 1 - H_q(\varrho + \eps_Z)}{\max} r\left(1 - \frac{\varrho}{H_q^{-1}(1 - r) - \eps_Z}\right)
\end{equation}
Here, $r$ and $R = 1 - \frac{\varrho}{H_q^{-1}(1-r) - \eps_Z}$ represent the rates of the random inner code and Folded Reed-Solomon code in the termination scheme respectively, and $\eps_Z \in (0,1)$ is a constant to be determined shortly. Note that the residual rate being below the Zyablov bound implies there are sufficient remaining channel uses to ensure the total number of transmissions does not exceed $n$, since we require the rate of the code used in the termination scheme to exceed Alice's residual message rate, which holds for our termination scheme since it can handle rates arbitrarily close to $R_Z$. As discussed in \hyperref[para:term_scheme]{Section 3.1.3}, if James has residual budget $1 - \frac{1}{q} - \gamma$ for some $\gamma \in (0, 1 - \frac{1}{q})$ when entering the termination scheme, the rate $R_Z$ is achievable using a list of size at most $(1/\gamma^2)^{\cO(\gamma^{-1}\log(1/\gamma))}$, implying that a constant number steps being required to reach termination retains this constant list size for our overall scheme. To bound the number of stages until this bound is reached, it suffices to lower bound the amount of descent in each stage, and calculate the number of stages required for a total descent of $1$ using this lower bound, since this guarantees the rate will be below the required rate for the termination scheme by this stage. Note that in each stage $i \geq 1$, the gap-to-capacity $\eps_i$ is an increasing function of $R_{i-1}$ due to \eqref{eq:eps_i_bound}. Thus, to bound the change in the gap-to-capacity between two adjacent stages, it suffices to derive a lower bound on $R_{i}$ for all $i \geq 1$.

The termination condition not being met at some arbitrary stage $i$ implies that $R_i(\varrho_i) \geq R_{\rm Z}(\varrho_i)$, meaning a lower bound on $R_{\rm Z}$ implies the same for $R_i$. With this in mind, we now proceed to derive a lower bound on $R_{\rm Z}$. Supposing James has residual budget $1 - \frac{1}{q} - \gamma$ when entering the termination scheme we have that
\begin{align}
    R_{\rm Z}\left(1 - \frac{1}{q} - \gamma\right) &\geq \left(1-H_q\left(1 - \frac{1}{q} - \frac{\gamma}{4}\right)\right) \nonumber \\
    \label{eq:zyablov_bound_1}&\hspace{58pt}\times\left(1 - \frac{1- \frac{1}{q}-\gamma}{1 - \frac{1}{q}- \frac{\gamma}{2}}\right) \\
     &=\left(1-H_q\left(1 - \frac{1}{q} - \frac{\gamma}4{}\right)\right) \nonumber \\
     \label{eq:zyablov_bound_2}& \hspace{60pt}\times\left(\frac{\gamma/2}{1-1/q-\gamma}\right) \\ 
    \label{eq:zyablov_bound_3} &\geq \left(1-H_q\left(1 - \frac{1}{q} - \frac{\gamma}{4}\right)\right)\frac{\gamma}{2} \\
    \label{eq:zyablov_bound_4} &\geq \frac{1}{2\ln q}\left(\frac{\gamma}{4}\right)^2\frac{\gamma}{2} \\
    \label{eq:zyablov_bound_5} &=\frac{\gamma^3}{64\ln q},
\end{align}
where:
\begin{itemize}
    \item \eqref{eq:zyablov_bound_1} follows from the choices\footnote{Note that the choice of $r$ satisfies the constraint in \eqref{eq:zyablov_def} and therefore provides a trivial lower bound on $R_Z$, since it involves a maximum over all feasible $r$.} $\eps_Z = \frac{\gamma}{4}$ and $r = 1 - H_q(1 - \frac{1}{q} - \frac{\gamma}{4})$ in \eqref{eq:zyablov_def};
    \item \eqref{eq:zyablov_bound_2} follows from combining terms and simplifying;
    \item \eqref{eq:zyablov_bound_3} follows since $0 < 1 - \frac{1}{q}- \frac{\gamma}{2} < 1$ for  $\gamma \in (0,1-\frac{1}{q})$;
    \item \eqref{eq:zyablov_bound_4} follows from the fact that $1 - H_q(\varrho) \geq \frac{1}{2\ln q}(1 - \frac{1}{q} - \varrho)^2$ for any $\varrho \in [0, 1-1/q]$, which can be seen by expressing $1 - H_q(\varrho)$ in terms of the KL divergence and applying Pinkser's Inequality.
\end{itemize}

Combining this bound on $R_Z$ with the assertion that $R_i \geq R_{\rm Z}$ for all stages $i$ before termination, it therefore follows that $R_i \geq \frac{\gamma^3}{64\ln q}$. From this, we are now ready to derive a bound on the number of stages until termination. Let $\tilde{\lambda}$ be the value such that 
\begin{equation}
    \frac{\eps_1}{\left(1-\frac{\gamma^3}{64\ln q}\right)^{\tilde{\lambda}}} = 1.
\end{equation}
Then, using \eqref{eq:eps_bound_final}, the fact that the change in gap-to-capacity in each stage is increasing in the rate, and the bound $R_j \geq \frac{\gamma^3}{64 \ln q}$ we have that
\begin{align}
    \label{eq:descent_1} \eps_{\tilde{\lambda}} &\geq \eps \prod_{j=1}^{\tilde{\lambda}-1}\frac{1}{1-R_j} \\
    \label{eq:descent_2} &\geq \frac{\eps}{\left(1 - \frac{\gamma^3}{64 \ln q}\right)^{\tilde{\lambda}-1}} \\
    \label{eq:descent_3} &=1,
\end{align}
which implies that $\tilde{\lambda}$ stages suffice to reach termination. Rearranging for $\tilde{\lambda}$ in \eqref{eq:descent_2}-\eqref{eq:descent_3} yields 
\begin{align}
    \label{eq:descent_4} \tilde{\lambda} &= \frac{\ln \eps}{\ln(1 - \frac{\gamma^3}{64\ln q})} + 1 \\
    \label{eq:descent_5} & =\frac{\ln\eps}{-\sum_{n=1}^{\infty}\frac{\gamma^{3n}}{(64\ln q)^nn}} + 1 \\
    \label{eq:descent_6} &\leq \frac{64(\ln q)\ln \frac{1}{\eps}}{\gamma^3} + 1 \\
    \label{eq:descent_7} &= \cO(\gamma^{-3}\log(1/\eps)),
\end{align}
where \eqref{eq:descent_5} follows from the Taylor series expansion of $\ln(1-x)$, and \eqref{eq:descent_6} follows by writing $\ln\eps = -\ln \frac{1}{\eps}$, canceling the negatives, and bounding the sum by its first term (since all terms in the sum are positive).

Letting $\lambda \coloneqq \min\{\ell \in \bN \mid R_{\ell}(\varrho_\ell) \leq R_{\rm Z}(\varrho_\ell)\}$ be the first stage where the rate lies below the Zyablov Bound, it follows that $\lambda \leq \tilde{\lambda}$, since $\epsilon_{\tilde{\lambda}} \geq 1$ due to \eqref{eq:descent_1}-\eqref{eq:descent_3}. Thus, since $\tilde{\lambda} = \cO(\gamma^{-3}\log(1/\eps))$ it follows that the same holds for $\lambda$, implying $\cO(\gamma^{-3}\log(1/\eps))$ stages suffice to reach the termination stage. To obtain the final result detailed in \Cref{thm:fullfb}, we derive a lower bound on $\gamma$. Due to \eqref{eq:eps_i_bound}, we have for any $\varrho' \in (0, 1 - \frac{1}{q})$ that $R_i(\varrho') \leq R_1(\varrho') = 1 - H_q(\varrho') - \eps$. Writing $\varrho' = 1 - \frac{1}{q} - \gamma$, and solving $R_1(\varrho') = 0$ we thus have that for all $i \geq 1$,
\begin{align}
    \label{eq:gamma_bound_1} R_i(\varrho') \leq R_1(\varrho') &=0 \\
    \label{eq:gamma_bound_2} \implies  \eps &= 1 - H_q\left(1 - \frac{1}{q} - \gamma \right)  \\
    \label{eq:gamma_bound_3} &= \frac{q^2\gamma^2}{2(q-1)\ln q} - o(\gamma^2)\\
    \label{eq:gamma_bound_4} &\leq \frac{q^2\gamma^2}{(q-1)\ln q},
\end{align}
where \eqref{eq:gamma_bound_3} uses the Taylor series expansion of $H_q(1 - 1/q - \gamma)$ centered at $1-1/q$ (see e.g., \cite[Eqn. 3.30]{guruswami2012essential}), and the fact that $\eps >0$ being arbitrarily small implies that $\gamma > 0$ is arbitrarily small due to the continuity of the $q$-ary entropy function. Rearranging for $\gamma$ then yields the bound $\gamma \geq \frac{1}{q}\sqrt{(q-1)\eps\ln q} = \Omega(\sqrt{\eps})$.
%
\subsection{Recursions with a penalty on the length of the next stage} 
\label{subsec:fullfb-imperfect-recursion}
As detailed in \Cref{subsec:fullfb-sync}, since Bob must guess error fractions $p_i$ on a $\delta$ grid, Alice too must round up her transmission lengths. Alice sends $\ell_{i+1}=\ell_i\parens*{H_q(\hp_i)+\eps_s}$, where $\eps_s=\Theta\parens*{\log(n)/n}$ is a slack to take care of type-counting -- it is henceforth ignored as it can be absorbed into other constants. Alice rounds up $\hp_i = \ceil{p_i/\delta}\cdot\delta$, hence $\hp_i-p_i\leq \delta$. The recursion does not follow \Cref{eqn:R_rho_update}, instead it satisfies :
\begin{align} 
&&
    R_{i+1} &= \frac{R_i H_q(\hp_i)}{1 - R_i} , & 
    \varrho_{i+1} &= \frac{\varrho_i - R_i p_i}{1 - R_i} . & 
& \notag 
\end{align}
Substituting the new value of $R_{i+1}$ into \eqref{eq:eps-recur}:
\begin{align}
    &\eps_{i+1} \geq \frac{\eps_i}{1-R_i} - \parens*{\frac{R_i}{1-R_i}}\parens*{H_q(\hp_i) - H_q(p_i)} \nonumber \\
    &\implies \eps_{i+1} \stackrel{(a)}{\geq} \frac{\eps_i-R_iH_q(\delta)}{1-R_i}=\frac{\eps_i-H_q(\delta)}{1-R_i} + H_q(\delta) \nonumber \\
    &\implies \eps_{i+1} - H_q(\delta)  \geq \frac{\eps_i-H_q(\delta)}{1-R_i}, \label{eq:app:rec:fullfb1}
\end{align}
where $(a)$  follows by noting that $H_q(\hp_i)-H_q(p_i) \leq H_q(\delta)$, since $\hp_i-p_i\leq \delta$. This can be seen by defining $F_1(x)=H_q(x-a)$ and $F_2(x)=H_q(x)-H_q(a)$, defined for $x\in[a,1]$. Consider $D(x) = F_1(x)-F_2(x)$, we have $D(a)=0$ and $D'(x)=F_1'(x)-F_2'(x)=H_q'(x-a)-H_q'(x) \geq 0$ since $H_q'(x)$ is a monotonically decreasing function. This implies $\abs*{H_q(b)-H_q(a)} \leq H_q(|b-a|)$ for all $a,b\in[0,1]$. Choosing a region where $H_q(\delta)$ is strictly increasing, i.e. $\delta\in[0,\frac{q-1}{q}]$, we have $\abs{b-a} \leq \delta \implies \abs{H_q(b)-H_q(a)} \leq H_q(\delta)$ for all $\delta\in[0,\frac{q-1}{q}]$, from which the result follows.

Denoting $H_q(\delta)=\Delta$ and unrolling the recursion \Cref{eq:app:rec:fullfb1} gives:
\begin{align}
    \eps_{i+1} - \Delta &\geq \frac{\eps_1 - \Delta}{\prod_{j=1}^{i} (1-R_j)} \nonumber \\
    \implies \eps_{i+1} &\geq \frac{\eps_1-i\parens*{H_q(\delta)}}{\prod_{j=1}^{i} (1-R_j)}, \label{eq:epsi-ineq-imperfect-recurrence}
\end{align}
by substituting $1\geq(1-R_k)$ in the numerator for all $k$ and defining $R_0\coloneq0$. Let $\tilde{\lambda}$ be a value such that 
\begin{equation}
    \frac{\eps_1/2}{\left(1-\frac{\gamma^3}{64\ln q}\right)^{\tilde{\lambda}-1}} = 1. \label{eq:imperfect-descent-1}
\end{equation}
Choosing $\delta$ such that $(\tilde{\lambda}-1)H_q(\delta)\leq\eps_1/2$ and substituting $i=\tlam-1$ into \Cref{eq:epsi-ineq-imperfect-recurrence} ensures
\begin{align*}
    \eps_{\tilde{\lambda}} &\geq \frac{\eps}{2} \prod_{j=1}^{\tilde{\lambda}-1}\frac{1}{1-R_j} \\
    &\geq \frac{\eps/2}{\left(1 - \frac{\gamma^3}{64 \ln q}\right)^{\tilde{\lambda}-1}} \\ 
    &=1.
\end{align*}
The extra factor of $\nicefrac{1}{2}$ in \Cref{eq:imperfect-descent-1} does not change the dependencies of $\tilde{\lambda}$ on $\gamma,\eps$ except by a constant factor, hence, from \Cref{eq:descent_7}, $\tilde{\lambda} = \cO(\gamma^{-3}\ln(1/\eps))$ holds. The bound on $\delta$ is then:
\begin{align*}
    &(\tilde{\lambda}-1)H_q(\delta) \leq {\eps/2} \\
    \implies &\delta \leq H_q^{-1}\parens*{\frac{\eps}{2(\tilde{\lambda}-1)}}\\
\end{align*}
Bounding the inverse entropy function as $H^{-1}_q(x) \lesssim c_q\frac{x}{\ln(1/x)}$, we can choose
\begin{equation*}
    \delta = \Theta\parens*{\frac{\eps/\tilde{\lambda}}{\ln(\nicefrac{\tilde{\lambda}}{\eps})}} = \Theta\parens*{\eps^{5/2}\ln^{-2}(1/\eps)},
\end{equation*}
by plugging in dependencies of $\tilde{\lambda}$ on $\gamma, \epsilon$, and noting that $\gamma = \Omega(\sqrt{\eps})$. 
The results in \Cref{thm:fullfb} follow.

\subsection{Recursions with a penalty on the length of the current stage}
\label{subsec:partfb-recursion}
As discussed in \Cref{subsec:partialfb-sync}, the length of the $i$-th stage is modified \emph{before} sending the next stage, as opposed to \Cref{subsec:fullfb-imperfect-recursion}, where the length of the current stage is fixed but alterations to the next stage are made. This produces a qualitatively different recursion as we shall see.

Recall that the length of the current stage, $\ell_i$, is rounded up (the new length denoted $\hell_i$) to the nearest multiple of $n\kappa$, in order to synchronize the start and end-point of stages between Alice and Bob. We thus have $\hell_i \coloneq \ceil*{\ell_i/n\kappa}\cdot n\kappa \implies \hell_i - \ell_i \leq n\kappa$. We define $\hR_i \coloneq \hell_i/n_i$, hence $\frac{1}{n_i}\paren*{\hell_i-\ell_i}=\hR_i-R_i\leq \frac{1}{n_i}n\kappa$. From discussions in \Cref{subsec:app:adv-ce-fullscheme-constants}, the length of the next stage is $\ell_{i+1}\leq\hell_i\parens*{H_q(p_i)+\eps_t}$. Dividing throughout by $n_{i+1}$ and noting that $n_{i+1} = n_i-\hell_i$, we have 
\begin{equation}
    R_{i+1}=\frac{\hell_i\parens*{H_q(p_i)+\eps_t}}{n_i-\hell_i} = \frac{\hR_i\parens*{H_q(p_i)+\eps_t}}{1-\hR_i}.
\end{equation}
With this expression for We define two notions of gap-to-capacity, 
\begin{equation}\label{eq:recur-eps}
    \eps_i \coloneq 1-H_q(\varrho_i)-R_i, \quad\quad\quad \heps_i \coloneq 1-H_q(\varrho_i)-\hR_i.
\end{equation}
Proceeding similarly to \Cref{subsec:fullfb-imperfect-recursion}, we can show
\begin{equation*}
    \eps_{i+1}=\frac{\heps_i-\eps_t}{1-\hR_i} + \eps_t, \quad\quad\quad \heps_i = \eps_i - (\hR_i-R_i).
\end{equation*}
This is a coupled recursion in $\eps_i$ and $\heps_i$, however the difficulty in solving this lies in bounding the quantity $\hR_i-R_i\leq n\kappa/n_i$. We reframe the recursion by defining $c_i\coloneq n_i/n$. We then have
\begin{align*}
    c_{i+1} = \frac{n_{i+1}}{n} = \frac{n_i-\hell_i}{n} &= \frac{(n_i-\ell_i)-(\hell_i-\ell_i)}{n}\\
    &\geq \frac{n_i(1-R_i)-n\kappa}{n}\\
    &\geq c_i\eps_i - \kappa.
\end{align*}
We now ``replace" the recursive variable $\heps_i$ with another one, $c_i$, to get a simultaneous recursion in terms of $\eps_i$ and $c_i$ instead. Absorbing the factor $\eps_t$ into the recursive variables by defining $u_i \coloneq \eps_i-\eps_t$ and $\hat{u}_i \coloneq \heps_i-\eps_t$ for ease of notation gives:
\begin{align*}
    (u_{i+1}) = \frac{\hat{u}_i}{1-\hR_i} &= \frac{u_i - (\hR_i-R_i)}{1-\hR_i} \\
    &\geq \frac{1}{1-\hR_i}\cdot \parens*{u_i - \frac{\kappa}{c_i}}, 
\end{align*}
from the definition of $c_i=n_i/n$ and the inequality $\hR_i-R_i\leq \kappa/c_i$.
Define $A_i\coloneq\frac{1}{1-\hR_i}\geq \frac{1}{1-R_\gamma} = A_\gamma\eqqcolon A$, where $R_\gamma = \frac{\gamma^3}{64\ln(q)}$ is the same lower bound we derived in \Cref{subsec:app:clean-descent}.
We thus again have a coupled recursion, but this time in terms of only $u_i$ and $c_i$:
\begin{equation}\label{eq:recurs-fullscheme}
    u_{i+1}\geq A_\gamma\parens*{u_i - \frac{\kappa}{c_i}}, \quad\quad c_{i+1} \geq c_i(u_i+\eps_t)-\kappa.
\end{equation}
We aim to keep, for all $i\leq\tilde{\lambda}$, $u_i\geq u_{\min}$ and $c_i \geq c_{\min}$. Intuitively, $u_i\geq u_{\min}$ keeps denominators $1-\hR_i$ bounded, and ensures each step still grows by a factor $A_i$, whereas $c_i$ bounded from below prevents the ``tax" on the clean recursion, $\kappa/c_i$, from becoming large. We will enforce both conditions with a single choice of parameter $\kappa$, then show that the scheme terminates successfully.

Unroll the recursion on $u_i$, denoting $\alpha_i\coloneq\kappa/c_i$:
\begin{equation*}
    u_i \hspace{-0.4pt}\geq \hspace{-0.4pt}A^{i-1}u_1 - A\sum_{j=1}^{i-1} A^{i-1-j}\alpha_j\hspace{-0.4pt} \geq \hspace{-0.4pt}A^{i-1}u_1 - S_i\alpha_{\max}(i)
\end{equation*}
where we have defined $\alpha_{\max}(i) = \max_{j<i} (\alpha_i)$ 
and $S_i\coloneq \sum_{j=1}^{i-1} A^{i-1-j}$. Set $i=\tilde{\lambda}$. We can write
\begin{align}
    \alpha_{\max} &< \frac{A^{\tilde{\lambda}-1}(u_1)-u_{\min}}{S_{\tlam}} \\
    \label{eq:k-from-u} \implies \kappa &\leq \frac{c_{\min}}{ S_{\tlam} } \parens*{A^{\tilde{\lambda}-1}u_1-u_{\min}},
\end{align}
the bound on $\kappa$ obtained by unrolling the recursion on $u_i$. Coming to the recursion on $c_i$, if we bound $u_i\geq u_{\min}$, we can write $c_{i+1} \geq c_ib-\kappa$, where $b=u_{\min}+\eps_t$. Solving this recursion gives:
\begin{equation*}
    c_i \geq b^{i-1} - \kappa \frac{b^{i-1}-1}{b-1}
\end{equation*}
A convenient sufficient target for all $i\leq \tlam$ is $c_{\min} \geq \frac{1}{2}b^{\tlam-1}$, which allows us to write another inequality on $\kappa$, this one being a result of unrolling the recursion on $c_i$:
\begin{equation}\label{eq:k-from-c}
    b^{\tlam-1} - \kappa\frac{b^{\tlam-1}-1}{b-1} \geq \frac{1}{2}b^{\tlam-1} \implies \kappa \leq \frac{1}{2}b^{\tlam-1}.
\end{equation}
We choose, for $\kappa$ to satisfy both \Cref{eq:k-from-u} and \Cref{eq:k-from-c}, $\kappa \leq M\cdot b^{\tlam}$, with the value of $M$ to be determined, and $b=u_{\min}+\eps_t$. Note that this makes
\begin{equation}\label{eq:kappa-by-cmin}
    \frac{\kappa}{c_{\min}} \leq \frac{Mb^{\tlam}}{\nicefrac{1}{2}b^{\tlam-1}} = 2Mb.
\end{equation}
Choose $u_{\min}=\eps/16$ and $\eps_t\leq\eps/32$, we then have $b=\Theta(\eps)$. Recall the recursive equations \Cref{eq:recurs-fullscheme}, where
\begin{equation*}
    u_{i+1}\geq A\parens*{u_i-\frac{\kappa}{c_i}} = Au_i\parens*{1-\beta_i},
\end{equation*}
where $\beta_i\coloneq\kappa/(c_iu_i)$. Intuitively, we want a stable multiplicative loss after unrolling this recursion: $\prod_{i=1}^{\tlam}(1-\beta_i)$ should be a constant, independent of $\lambda$. We have
\begin{equation}\label{eq:beta-value}
    \beta_i = \frac{\kappa}{c_iu_i}\leq \frac{2M}{u_{\min}} b^{\tlam-(i-1)},
\end{equation}
so $\beta_i$ forms a ``reversed" geometric series - the last few stages contribute the most. Enforcing $\beta_i\leq 1/2$, we can use the inequaltiy $1-x \geq e^{-2x}$ for all $x\in(0,1/2)$ to bound the multiplicative loss:
\begin{equation*}
    \prod_{i=1}^{\tlam} (1-\beta_i) \geq \exp\parens*{-2\sum_{i=1}^{\tlam}  \beta_i}
\end{equation*}
Using the bound on $\beta_i$ from \Cref{eq:beta-value}, 
\begin{align*}
    \sum_{i=1}^{\tlam}\beta_i \leq \frac{2M}{u_{\min}}\sum_{t=1}^{\tlam} b^t \leq \frac{2M}{u_{\min}}\cdot\frac{b}{1-b} &\leq \frac{32M}{\eps}\cdot\frac{\eps}{15} \\
    &= \frac{32}{15}M.
\end{align*}
This is to demonstrate that our multiplicative ``loss" is small. and the factor $A^{\tlam}$ in the unrolled recursion for $u_i$ ensures that $\eps_{\tlam}$ can be driven up to reach 1, ensuring we terminate. Note that we enforced $\beta_i\leq 1/2 \;\forall i$, to make sure this is true we must have $\dfrac{\kappa}{c_{\min}} \leq \dfrac{u_{\min}}{2}$. Since we already have, from \Cref{eq:kappa-by-cmin}, $\kappa/c_{\min} \leq 2Mb$, choosing $M \leq 1/16$ ensures the condition is satisfied. $M$ must also satisfy \Cref{eq:k-from-u}, we have
\begin{align*}
    \frac{\kappa}{c_{\min}} \leq 2Mb &\leq \frac{1}{S_{\tlam}}\cdot \parens*{A^{\tlam-1}u_1-u_{\min}} \\
    \implies M &\leq \frac{1}{2bS_{\tlam}}\cdot \parens*{A^{\tlam-1}u_1 - u_{\min}}\\
    \implies M &\lesssim \frac{A-1}{2A}\cdot\frac{u_1}{b} = \cO(A-1) = \cO(R_{\gamma}) \\
    &= \cO(\gamma^3).
\end{align*}
We have that $\gamma=\Theta(\sqrt{\eps})$, hence choosing $M =\Theta(\eps^{3/2})$ satisfies all required inequalities.
Recall we chose $\kappa=Mb^{\tlam} = \Theta(\eps^{\tlam+\nicefrac{3}{2}})$, giving the desired result.
\section{Analysing the performance of various protocols}

For a guide to the constants used throughout all schemes, see \Cref{tab:constants}. We attempt to keep subscripts in line with the part of the scheme that the constants relate to. For example, Bob sends $C_e\log(N)$ symbols for Alice to estimate the noise level $p$ - here the subscript `e' stands for ``estimate". The table assumes that the stage in question has length $N$, and James corrupts a fraction $p$ of symbols in that stage.

\begin{table}[t]\label{tab:constants}
  \centering
  \caption{Guide to constants used throughout the paper.}
  \label{tab:constants}
  \setlength{\tabcolsep}{8pt}
  \begin{tabularx}{\linewidth}{@{} l Y @{}}
    \toprule
    \textbf{Constant} & \textbf{Meaning / Role} \\
    \midrule
    \constrow{\eps}{The rate-slack to information theoretic capacity, $R=1-H_q(\varrho)-\eps$}
    \constrow{C_e}{Alice \underline{e}stimates $p$ with $C_e\log(N)$ feedback samples}
    \constrow{\eps_e}{Threshold above which Alice's \underline{e}stimate $\tilde{p}$ is considered ``bad"}
    \constrow{C_c}{Alice divides a stage into \underline{c}hunks of length $C_c\log(N)$}
    \constrow{C_p}{Bob sends $C_p\log(N)$ symbols as feedback for Alice to select her \underline{p}ermutation}
    \constrow{\eps_p}{The fraction of ``bad'' \underline{p}ermutations in $\cP$ is desired to be at most $ N^{-\eps_p} $}
    \constrow{\eps_T}{Threshold for empirical \underline{t}ype of permuted chunk to deviate from $p$}
    \constrow{\eps_N}{Allowed fraction of bad chunks under a single permutaiton}
    \constrow{\eps_d}{Slack for Bob's typical set check}
    \constrow{\eps_h}{Slack for Alice's \underline{h}ash function output}
    \constrow{\delta}{The discretization level of Bob's guesses, i.e. $\hp_i=k\delta$ for some $k\in\bZ^+$}
    \constrow{\tlam}{An upper bound on the number of stages until termination}
    \constrow{\eps_t}{(In the final scheme) ``\underline{T}otal" slack to upper bound the length of the next stage}
    \constrow{\kappa}{(In the final scheme) Bob sends feedback after every $n\kappa$ forward transmissions}
    \bottomrule
  \end{tabularx}
\end{table}

\subsection{\Cref{prot:sw_no-eff}}
\label{subsec:app:sw-no-ce-proof}

The error event is one where there exists a vector typical with $\vbfy$ and also hashed to the same vector as $\vbfx$; or, that $\vbfx$ is not itself typical.
\begin{align}
    \Pr\paren{h(\iterx) = h(\vbfx)} &= \dfrac{1}{2^{N\cdot H(\bfx|\bfy)+\eps_h}}, \\
    \hspace{100pt}&\forall \; \iterx : (\vbfy, \iterx) \in \cT^{(N)}_{\bfy,\bfx}, \; \; \iterx \neq \vbfx \nonumber.
\end{align}
Taking a union bound over all $\iterx \in \cT^{(N)}_{\bfx|\bfy}(\eps_d)$, where $\abs{\cT^{(N)}_{\bfx|\bfy}} \leq 2^{N\cdot(H(\bfx|\bfy)+\eps_d)}$, we have that for sufficiently large $N$,
\begin{equation}
    \label{eq:err-bound_single-chunk}
    \Pr\paren{\hvbfx \neq \vbfx} \leq 2^{-N(\eps_h-\eps_d)} + e^{-N\eps_d^2} \leq 2^{-N\eps_h/3}.
\end{equation}

\subsection{\Cref{prot:no-adv_ce}}
\label{subsec:app:no-adv-ce-proof}

\subsubsection{Parameter Selection}
1. Output vector length of R.S Codes: Due to the hashing scheme, each chunk has a probability of being decoded in error, $P_e^{(i)}=N^{-C_c\eps_H}$. To combat this we choose
\begin{equation}
    K' = \frac{K}{1-4N^{-C_c\eps_H}},
\end{equation}
this choice of $K'$ is shown to be sufficient in the immediately following section.

\noindent 2. Field size of Reed Solomon Codes: We define our $(K',K)$ RS code over $\bF_Q$ with $Q=q^{\lchn}=N^{C_c}$, hence we must have $K'\leq Q-1$, giving us
\begin{equation*}
    K'=\frac{N}{C_c\log(N)\cdot(1-4N^{-C_c\eps_H})} \leq N^{C_c}.
\end{equation*}
This inequality is satisfied for any value of $C_c\geq 1$.

3. Hashing slack parameters: Identical to the working in \Cref{subsec:app:sw-no-ce-proof}, the scheme works if $\eps_h>\eps_d$. Concretely we pick $\eps_h>0$ independent of $N$ and $\eps_d=\eps_h/2$. We denote $\eps_H=\eps_h/3$ merely for notational brevity.

\subsubsection{Error probability}
\begin{proof}
    From \Cref{eq:pei} the probability that each chunk is in error is $P_e^{(i)}=N^{-C_c\eps_H}$. Let $\bfn_e$ denote the number of chunks in error. Then, in our $(K',K)$ RS code, $\bE[\bfn_e] = K'\cdot P_e^{(i)}$. Because the hash functions are independent, the error events are also independent and we can apply the Chernoff bound,
    
    \begin{align}
        \Pr\paren{\bfn_e > 2\bE[\bfn_e]} &\leq e^{-\frac{1}{3}K'\cdot P_e^{(i)}} \\
        \label{eq:no-adv-ce_chunk-err}&\hspace{-20pt}= \exp\parens*{-\frac{1}{3}\cdot\frac{N^{1-C_c\eps_H}}{C_c\log N\paren{1-4N^{-C_c\eps_H}}}}.
    \end{align}
    
    Now, the number of errors correctable by the proposed $(K',K)$ code is

    \begin{align*}
        t = \frac{K'-K}{2} &= \frac{1}{2}\cdot\paren{K' - K'\paren{1-4N^{-C_c\eps_H}}} \\
        &= K'\cdot 2N^{-C_c\eps_H} = 2\bE[\bfn_e]
    \end{align*}
    
    The scheme only fails when $\bfn_e > t = 2\bE[\bfn_e]$, which happens with probability given by \Cref{eq:no-adv-ce_chunk-err}, which is the required result.
\end{proof}

\begin{remark}
    The cost overhead for storing all the hash functions is  $\cO(\poly(N))$. Each hash function $h_i$,  $\forall \;i \in [K']$, requires $\abs{\cT^{(\lchn)}_{\eps_d}({\bfy})}\cdot \abs{\alphX^{\lchn(H(\bfx|\bfy)+\eps_h)}}$ symbols to specify (let $\lchn'\coloneqq \lchn ( H(\bfx|\bfy)+\eps_h)$ ), making the total (offline) storage cost:
    \begin{equation}
        \abs{\cT^{(\lchn)}_{\eps_d}{(\bfx)}}\cdot \abs{\alphX^{\lchn'}} \cdot K' \approx q^{\lchn H(\bfx)}\cdot q^{\lchn'} \cdot K' \leq \frac{N^{2C_c}}{C_c\log(N)}.
    \end{equation}
Choosing $C_c=1$ caps this at $N^2$ upto constant factors.
\end{remark}

\subsection{\Cref{prot:adv-ce}}
\label{subsec:app:adv-ce-proof}

\subsubsection{Parameter selection}
\label{subsubsec:app:adv-ce-params}

1. \uline{Output vector length of R.S Codes}: Due to both the hashing scheme and the selected permutation potentially causing errors in the decoding of the chunks, we choose 
\begin{equation}\label{eq:adv-ce:kdash}
    K' = \frac{K}{1-(4q^{-\sqlch/2} + 2\eps_N)},
\end{equation}
where $\eps_N$ is an upper bound on the fraction of chunks that are \emph{not} quasi-uniform with respect to James' true error fraction on the stage, $p_i$ (see \Cref{sec:perm} for details). This choice of $K'$ will be shown to be apt in the immediately following ``Error probability" section.

\noindent 2. \underline{Choice of Hash slack and Bob's Typical Set}: There are two factors that confuse Alice's estimate of the error fraction $\tilde{p}$, from the true error fraction in the permuted chunk. Note the error fraction used by James in the entire \emph{stage} is denoted $p$. For chunk $i$, let $t_i = \frac{1}{\lchn}\cdot \dh\parens*{\permed_{(i)}, \ypermed_{(i)}}$, the actual error fraction in the permuted chunk. We know, for a \emph{good} permutation chunk, $|t_i - p|\leq \eps_T$ (see \Cref{sec:perm}), and for a good estimate $\tilde{p}$, we have $|\tilde{p}-p|\leq\eps_e$. Combining, we have $|t_i-\tilde{p}|\leq \eps_T+\eps_e$. Along with a tiny, $o(1)$ slack, choosing $\eps_d$ (the slack parameter for Bobs typical set check) such that $\eps_d \geq \eps_T + \eps_e + 2/\lchn$ ensures that $t_i\cdot\lchn \in [\lchn(\tilde{p}-\eps_d), \lchn(\tilde{p}+\eps_d)]$. Choosing $\eps_h > \eps_d$ ensures successful execution of the protocol, in specific we choose $\eps_h = 2\eps_d$.

\noindent 3. \underline{Permutation slack parameters}: We would like to have $\Pr(|t_i-p|>\eps_T)\leq c\eps_N$, (for some constant c), this ensures that the probability that the number of ``bad" chunks produced by our permutation will be $\leq K'\cdot \eps_N$ w.h.p, ensuring that the R.S. code with parameter $(K',K)$ will correct those erroneous chunks. Choosing $\eps_T = \sqrt{\frac{6\log(\lchn)}{\lchn}}$ ensures $\Pr(|t_i-p|>\eps_T) \leq 2\exp(-2\eps_T^2\lchn) \ll c\eps_N$. From \Cref{lem:pgood}, to keep the exponent positive, we must choose $1+C_p-\eps_p > 0$ - simply choosing any $C_p>\eps_p$ will suffice.

\subsubsection{Error probability}
The following are error events for \Cref{prot:adv-ce}:
\begin{enumerate}
    \item Event $A$: A permutation $\pi\in\cP$ is ``bad'' if for some choice of $ \vs $ by James, upon permuted by $\pi$, the resulting sequence does not have a sufficiently large fraction of chunks whose types are sufficiently close to the type of $\vs$. 
    The permutation bank $\cP$ is ``bad'' if there are more than a fraction $N^{-\eps_p}$ of bad permutations in $\cP$. See \Cref{sec:perm} for formal definitions. 
    By \Cref{lem:perm_param}:
    \begin{align}
        \Pr(A) &\leq \exp_2\parens*{-\frac{N^2}{64q^2\lchn^8}} . \notag 
    \end{align}
    \item Event $B$: Even conditioning on a good permutation bank $\cP$ (in the sense that $ A^c $ holds), a bad permutation in $\cP$ may still exist and be chosen by Alice under her uniform sampling. 
    Also, James may, by sheer chance, guess Alice's permutation (even if it is a good one). 
    The probability of at least one of these happening is at most 
    \begin{align}
        \Pr(B) &\le N^{-\eps_p} + N^{-C_p} \le 2 \cdot N^{-\eps_p}. \notag 
    \end{align}
    \item Event $C$: Alice's estimate of the error fraction $\wt{p}$ poorly approximates the true frequency $p$ induced by James in the sense that $ \abs{\wt{p} - p} > \eps_e $. 
    Invoking \Cref{thm:learning-distr-linfty}, as discussed in \hyperref[para:est-nl]{Section 4.2.1}, the probability of this happening is at most
    \begin{align}
        \prob{C} &\le 2 \cdot e^{-\frac{1}{2} \abs{T} \eps_e^2} = 2\cdot N^{-\frac{1}{2\ln(q)}C_e\eps_e^2}, \notag 
    \end{align}
    where $ \abs{T} = C_e\log(N) $ is given in \Cref{prot:adv-ce}.
    \item Event D: The R.S outer code sees too many bad chunks. Conditioning on a good permutation bank $\cP$, a good permutation $\pi$ and a good estimate $\tilde{p}$, the results in \Cref{sec:perm} ensure that the fraction of bad chunks is $< \eps_N$ with small probability. Thus setting $K'$ to the value in \Cref{eq:adv-ce:kdash} ensures we can handle the hash errors and permutation errors with probability of failure dominated by
    \begin{equation*}
        \exp\parens*{-\Theta\parens*{\frac{N}{\lchn}q^{-\sqlch/2}}}
    \end{equation*}
\end{enumerate}
Combining the error events, we see that the error probability of the scheme is dominated by events $B$ and $C$, i.e.
\begin{equation}\label{eq:error-prob-full}
    P_e(N) \leq 2\parens*{N^{-\eps_p}+N^{-C_e\eps_e^2}}.
\end{equation}
Union bounding over a constant $\tlam$ number of stages retains the asymptotic dependence for large enough $N$ (and hence $n$). 
\begin{remark}
\label{rem:p_error_discussion}
    Event $B$ includes the possibility that Alice selects a ``bad'' permutation, or that James happens to guess the permutation she selected, while Event $C$ corresponds to an error in Alice estimating the noise level James used, from the feedback she receives from Bob. These probabilities are inverse-polynomial in $n$ because the choice of permutation is made from a polynomial-sized bank, while the sampling used for estimating the noise level involves only \(O(\log n)\) observations. To obtain exponentially small error, such as \( e^{-\Omega(n)} \), these events themselves would need to have exponentially small probability. With the present approach, this would require much larger random objects or substantially more sampling; for example, a permutation bank whose size grows exponentially with \(n\) (requiring exponential storage), or feedback samples whose size grows linearly rather than logarithmically with \(n\) - resulting in a non-vanishing feedback rate.
\end{remark}

\subsection{The full scheme with partial feedback}
\label{subsec:app:adv-ce-fullscheme-constants}

Here we detail some minor variations in protocol parameter selection for the complete Slepian-Wolf/Weldon-type recursive scheme, as compared to the toy example - whose parameters choices are explained in (\Cref{subsubsec:app:adv-ce-params}). 

The choice of hash slacks and Bob's typical set must be altered. Since, in the recursive Weldon scheme, Alice further rounds up her estimate in stage $i$, $\tilde{p}_i$, to $\hp = \ceil{\tilde{p}_i/\delta}\cdot\delta$, Bob's typical set slack must satisfy $\eps_d \geq \delta + \eps_T + \eps_e$. Choosing $\eps_e = \delta/2$ (providing an explicit choice) and $\eps_d = 2\delta$ ensures this, since $\eps_T = o(1)$. All other parameters stay the same. We can hence upper bound the length of the next stage by $\ell_{i+1}\leq\hell_i\parens*{H_q(p_i)+\eps_t}$, by choosing $\eps_t>\eps_d$.

The error probability of the scheme, union bounded over the (constant) number of stages, is asymptotically the same as \Cref{eq:error-prob-full}. Setting the error probability of the scheme in accordance with our design parameter $\eta$, we have $\eta = \min\{\eps_p, C_e\eps_e^2\}$. Since $C_P>\eps_p>\eta$, and the size of the permutation bank we store is $n^{C_p}$, this bottlenecks the tradeoff between error probability and offline storage costs. 
Since each permutaiton needs $\cO(n\log(n))$ storage space, this results in a required total budget of $\cO(n^{\eta+1}\log n)$
Note that $C_e\eps_e^2$ can be made large (upto constant factors) since $C_e$ dictates the size of feedback Bob sends for Alice to estimate $p$, which is $C_e\log^2(n) = o(n)$ feedback - there are no asymptotic storage tradeoffs here. Hence the probaility of error is $\cO(n^{-\eta})$ and the storage costs are $\cO(n^{\eta+1}\log n)$ for an appropriately chosen $\eta>1$, proving the required performance metrics.
\section{A lemma regarding concentration over random permutations}
\label{sec:perm}

Denote by $ \allperms $ the symmetric group of order $n$, i.e., the set of all permutations on $n$ elements. 
Let $ \cP = \braces*{\perm^1, \cdots, \perm^P} \subset \allperms $ be a set of permutations where $ P = n^{C_p} $ for a constant $ C_p>0 $. 
Let $ \vs\in\cS^n $, $ \perm\in \allperms $ and $ i\in[N] $. 
For $ \slack{T} \in (0,1) $, define 
\begin{align}
    \cE_{\ref{lem:pgood}}(\vs, \perm, i) &\defeq
    \braces*{
        \norm*{\type{\perm(\vs)_{\cI_i}} - \type{\vs}}_\infty \le \slack{T}
    } . \label{eqn:def_E_pgood_s_pi_i} 
\end{align}
For $ \slack{N} \defeq 1/\ell^3 $, define 
\begin{align}
    \cE_{\ref{lem:pgood}}(\vs, \perm) &\defeq 
    \braces*{
        \frac{1}{N} \sum_{i = 1}^N \one_{\cE_{\ref{lem:pgood}}(\vs, \perm, i)} \ge 1 - \slack{N}
    } . \notag 
\end{align}
For $ \eps_p > 0 $, define 
\begin{align}\label{eq:perms:good-pi}
    \cE_{\ref{lem:pgood}}(\vs) &\defeq \braces*{ \frac{1}{P} \sum_{j = 1}^P \one_{\cE_{\ref{lem:pgood}}(\vs, \perm^j)} \ge 1 - n^{-\eps_p} }  .
\end{align}
Finally, let
\begin{align}
    \cE_{\ref{lem:pgood}} \defeq \bigcap_{\vs\in\cS^n} \cE_{\ref{lem:pgood}}(\vs) . \notag 
\end{align}

We will choose 
\begin{align}
    \cS = \Brackets{q}, \
    \ell = \lchn, \
    \eps_T = \sqrt{ \frac{6\log(\lchn)}{\lchn} }, \
    C_p = 2, \
    \eps_p = 1 \label{eqn:choose_param} 
\end{align}
and prove the following estimate for the probability of the undesirable event $\cE_{\ref{lem:pgood}}^c$ when each element in $\cP$ is drawn independently and uniformly from $\allperms$. 

\begin{lemma}
\label{lem:perm_param}
Let $\cP = \braces*{ \rperm^1, \rperm^2, \ldots, \rperm^P }$, and suppose that $ \cP \sim \unif(\allperms^P)$. 
Then for all sufficiently large $n$, 
\begin{align}
    \prob{ \cE_{\ref{lem:pgood}}^c } &\le \exp_2\parens*{ - \frac{ n^{2}}{64 q^2 \lchn^8} } . \notag 
\end{align}
\end{lemma}

\Cref{lem:perm_param} follows from a sequence of lemmas below. 

\begin{lemma}
\label{lem:pgood1}
Let $ \rperm\sim\unif(\allperms) $. 
Then for any $ \vs\in\cS^n $ and every sufficiently large $n$, 
\begin{align}
    \prob{ \cE_{\ref{lem:pgood}}(\vs, \rperm)^c } &\le 2\abs*{\cS} \exp\parens*{ -\frac{\slack{N}^2 n}{16 \abs*{\cS}^2 \ell^2} } . \notag 
\end{align}
\end{lemma}

\begin{proof}
We start by writing the event $ \cE_{\ref{lem:pgood}}(\vs, \rperm, i)^c $ as: 
\begin{align}
    \cE_{\ref{lem:pgood}}(\vs, \rperm, i)^c
    &= \bigcup_{s\in\cS} \cF(\vs, \rperm, i, s) , 
    \label{eqn:decomp_E} 
\end{align}
where 
\begin{align}
    \cF(\vs, \rperm, i, s) &\defeq \braces*{ \abs*{ \type{\rperm(\vs)_{\cI_i}}(s) - \type{\vs}(s) } > \slack{T} } . 
\label{eqn:F_+-}
\end{align}

With the above notation, we upper bound the probability of $ \cE_{\ref{lem:pgood}}(\vs, \perm)^c $ as follows:
\begin{align}
    \prob{\cE_{\ref{lem:pgood}}(\vs, \rperm)^c}
    &= \prob{ \frac{1}{N} \sum_{i = 1}^N \one_{\cE_{\ref{lem:pgood}}(\vs, \rperm, i)^c} \ge \slack{N} } \notag \\
    &\le \prob{ \frac{1}{N} \sum_{s\in\cS} \sum_{i = 1}^N \one_{\cF(\vs, \rperm, i, s)} \ge \slack{N} } \label{eqn:one} \\
    &\hspace{-5.71pt}\le \prob{ \bigcup_{s\in\cS} \braces*{ \frac{1}{N} \sum_{i = 1}^N \one_{\cF(\vs, \rperm, i, s)} \ge \frac{\slack{N}}{\abs*{\cS}} } } \label{eqn:event} \\
    &\le \sum_{s\in\cS} \prob{ \frac{1}{N} \sum_{i = 1}^N \one_{\cF(\vs, \rperm, i, s)} \ge \frac{\slack{N}}{\abs*{\cS}} }
    . \label{eqn:term12} 
\end{align}
In \eqref{eqn:one}, we use the fact that $ \one_{\cE_1\cup\cE_2} \le \one_{\cE_1} + \one_{\cE_2} $ for any two events $ \cE_1, \cE_2 $. 
The inequality \eqref{eqn:event} holds since the event in \eqref{eqn:one} implies that in \eqref{eqn:event}. 
To see so, consider the complement of the event in \eqref{eqn:event}:
\begin{align}
    \bigcap_{s\in\cS} \braces*{ \frac{1}{N} \sum_{i = 1}^N \one_{\cF(\vs, \rperm, i, s)} < \frac{\slack{N}}{\abs*{\cS}} } , \notag 
\end{align}
which implies
\begin{align}
    \frac{1}{N} \sum_{s\in\cS} \sum_{i = 1}^N \one_{\cF(\vs, \rperm, i, s)} 
    &< \abs*{\cS} \cdot \frac{\slack{N}}{\abs*{\cS}} = \slack{N} , \notag 
\end{align}
contradicting the event in \eqref{eqn:one}. 

It now remains to bound the probability in \eqref{eqn:term12}. 
We will use the following result due to Maurey \cite{Maurey} with a subsequent improvement by Schechtman \cite{Schechtman}. 
The precise statement below is taken from \cite[Theorem 1]{Schechtman}. 
Let $ f \colon \allperms \to \bR $ be an $L$-Lipschitz function with respect to the normalized Hamming metric, that is, 
\begin{align}
  \abs*{ f(\perm) - f(\sigma) } &\le L \frac{\dh(\perm, \sigma)}{n} , \qquad 
  \forall \perm, \sigma \in \allperms , \label{eqn:Lip} 
\end{align}
where 
\begin{align}
  \dh(\perm, \sigma) &= \sum_{i = 1}^n \indicator{\perm(i) \ne \sigma(i)} . \notag 
\end{align}
Then it holds for any $ t>0 $ that
\begin{align}
   \prob{ \abs*{ f(\rperm) - \expect*{f(\rperm)} } \ge t }
   &\le 2 \exp\parens*{ -\frac{t^2n}{4 L^2} } , \label{eqn:Maurey} 
\end{align} 
where the randomness comes from $ \rperm \sim \unif(\allperms) $. 
In our case of \eqref{eqn:term12}, 
\begin{align}
  f(\perm) &= \frac{1}{N} \sum_{i = 1}^N \one_{\cF(\vs, \perm, i, s)} . \notag 
\end{align}
Let us examine the Lipschitz coefficient of $ f $. 
Consider $ \perm, \sigma \in \allperms $ with $ \dh(\perm, \sigma) = d $. 
Then $ \cF(\vs, \perm, i, s) $ and $ \cF(\vs, \sigma, i, s) $ differ for at most $d$ values of $i$, implying $ \abs*{f(\perm) - f(\sigma)} \le d/N $. 
Therefore $ f $ is $L$-Lipschitz in the sense of \eqref{eqn:Lip} with $ L = n/N = \ell $. 
Applying \eqref{eqn:Maurey} gives
\begin{align}
  &\prob{ \frac{1}{N} \sum_{i = 1}^N \one_{\cF(\vs, \rperm, i, s)} \ge \frac{\slack{N}}{\abs*{\cS}} } \\
  &= \bP \Bigg(\frac{1}{N} \sum_{i = 1}^N \one_{\cF(\vs, \rperm, i, s)} - \expect*{\one_{\cF(\vs, \rperm, 1, s)}} \\
  &\hspace{113pt}\ge \frac{\slack{N}}{\abs*{\cS}} - \expect*{\one_{\cF(\vs, \rperm, 1, s)}}\Bigg) \notag \\
  &\le 2 \exp\parens*{ - \frac{\parens*{ \slack{N}/\abs*{\cS} - \expect*{\one_{\cF(\vs, \rperm, 1, s)}} }^2 n}{4 \ell^2} } \notag \\
  &\le 2 \exp\parens*{ -\frac{\parens*{\slack{N}/(2\abs*{\cS})}^2 n}{4 \ell^2} } \label{eqn:large_n} \\
  &= 2 \exp\parens*{ -\frac{\slack{N}^2 n}{16 \abs*{\cS}^2 \ell^2} } . \notag 
\end{align}
To obtain \eqref{eqn:large_n}, we use Lemma \ref{lem:mean_ub} below to upper bound $ \expect*{\one_{\cF_+(\vs, \rperm, 1, s)}} $ and note also that the assumption $ \eps_N = 1/\ell^3 $ implies $ \eps_N/\abs*{\cS} - 2e^{-2\eps_T^2 \ell} \ge \eps_N/(2\abs*{\cS}) $ for every sufficiently large $n$. 
\end{proof}

\begin{lemma}
\label{lem:mean_ub}
Let $ \rperm \sim \unif(\allperms) $. 
Then for any $ i\in[N] $, $ \vs\in\cS^n $ and $ s\in\supp(\type{\vs}) $, 
\begin{align}
  \prob{ \cF(\vs, \rperm, i, s) } &\le 2e^{-2\eps_T^2 \ell} . 
  \label{eqn:F+-_ineq} 
\end{align}
\end{lemma}

\begin{proof}
Recall from \eqref{eqn:F_+-} that 
\begin{align}
  \prob{ \cF(\vs, \rperm, i, s) }
  &= \bP\Bigg(\abs*{ \frac{1}{\ell} \sum_{k\in\cI_i} \indicator{ \rperm(\vs)_k = s } - \type{\vs}(s) } \notag \\
  &\hspace{120pt}> \eps_T \Bigg) . \notag 
\end{align}
Note that 
\begin{align}
  \expect*{ \frac{1}{\ell} \sum_{k\in\cI_i} \indicator{ \rperm(\vs)_k = s } }
  &= \prob{ \vs_{\rperm(1)} = s }
  = \type{\vs}(s) , \notag 
\end{align}
and for any $m$, 
\begin{align}
  \prob{ \sum_{k\in\cI_i} \indicator{ \rperm(\vs)_k = s } = m }
  &= \frac{\binom{n \type{\vs}(s)}{m} \binom{n - n \type{\vs}(s)}{\ell - m}}{\binom{n}{\ell}} . \notag 
\end{align}
Therefore, 
\begin{align}
  \sum_{k\in\cI_i} \indicator{ \rperm(\vs)_k = s } &\sim \hypergeom(n, n \type{\vs}(s), \ell) , \notag 
\end{align}
where $ \hypergeom(n, m, \ell) $ denotes the hypergeometric distribution with population size $n$, number of successes $m$ and number of draws $\ell$. 
The standard Hoeffding's inequality for hypergeometric distribution then guarantees the validity of \eqref{eqn:F+-_ineq}. 
\end{proof}

\begin{lemma}
\label{lem:pgood2}
Let $\cP = \braces*{ \rperm^1, \rperm^2, \ldots, \rperm^P }$, and suppose that $ \cP \sim \unif(\allperms^P)$. Then for any $ \vs\in\cS^n $ and every sufficiently large $n$, 
\begin{align}
    \prob{ \cE_{\ref{lem:pgood}}(\vs)^c }
    &\le \exp_2\parens*{ - \frac{ n^{1 + C_p - \eps_p}}{32 \abs*{\cS}^2 \ell^8} } . \notag 
\end{align}
\end{lemma}

\begin{proof}
Since all permutations in $\cP$ are independent and uniformly distributed over $\allperms$, 
\begin{align}
    &\prob{ \cE_{\ref{lem:pgood}}(\vs)^c }
    = \prob{ \frac{1}{P} \sum_{j = 1}^P \one_{\cE_{\ref{lem:pgood}}(\vs, \perm^j)^c} \ge n^{-\eps_p} } \notag \\
    &= \sum_{k = n^{-\eps_p} P}^P \binom{P}{k} \expect*{\one_{\cE_{\ref{lem:pgood}}(\vs, \perm^1)^c}}^{k} \notag \\
    &\hspace{100pt}\times \parens*{1 - \expect*{\one_{\cE_{\ref{lem:pgood}}(\vs, \perm^1)^c}}}^{P - k} \notag \\
    &\le P \binom{P}{n^{-\eps_p} P} \prob{\cE_{\ref{lem:pgood}}(\vs, \perm^1)^c}^{n^{-\eps_p} P} \label{eqn:dom} \\
    &\le P \cdot P^{n^{-\eps_p} P} \cdot \parens*{ 2 \abs*{\cS} \exp\parens*{ -\frac{\slack{N}^2 n}{16 \abs*{\cS}^2 \ell^2} } }^{n^{-\eps_p} P} \notag \\
    &= \exp_2\bigg(\hspace{-3pt}\log(P) + n^{-\eps_p} P \log(P) + n^{-\eps_p} P \log(2\abs*{\cS}) \notag \\
    &\hspace{134pt}- n^{-\eps_p} P \frac{\slack{N}^2 n}{16 \abs*{\cS}^2 \ell^2} \bigg) \notag \\
    &= \exp_2\bigg( C_p\log(n) + C_p n^{C_p - \eps_p} \log(n)  \notag \\
   &\hspace{60pt}+ n^{C_p - \eps_p} \log(2\abs*{\cS}) - \frac{ \slack{N}^2 n^{1 + C_p - \eps_p}}{16 \abs*{\cS}^2 \ell^2} \bigg) \notag \\
    &\le \exp_2\parens*{ - \frac{ \slack{N}^2 n^{1 + C_p - \eps_p}}{32 \abs*{\cS}^2 \ell^2} } . \notag 
\end{align}
The inequality \eqref{eqn:dom} follows since the largest term in the sum is the first one. 
The last inequality holds for all sufficiently large $n$. 
\end{proof}

Finally the following lemma follows immediately from Lemma \ref{lem:pgood2} by taking a union bound over $ \vs\in\cS^n $. 

\begin{lemma}
\label{lem:pgood}
Let $\cP = \braces*{ \rperm^1, \rperm^2, \ldots, \rperm^P }$, and suppose that $\cP \sim \unif(\allperms^P)$. 
Then for all sufficiently large $n$, 
\begin{align}
    \prob{ \cE_{\ref{lem:pgood}}^c } &\le \exp_2\parens*{ - \frac{ \slack{N}^2 n^{1 + C_p - \eps_p}}{64 \abs*{\cS}^2 \ell^2} } . \notag 
\end{align}
\end{lemma}

Specializing \Cref{lem:pgood} to the configuration of parameters in \Cref{eqn:choose_param} yields the promised \Cref{lem:perm_param}. 

\section{A lemma regarding type-class estimation from i.i.d.\ samples}\label{sec:est-noise}


The following theorem is a direct consequence of the Dvoretzky--Kiefer--Wolfowitz inequality \cite{Dvoretzky_Kiefer_Wolfowitz} (with the sharp constant $2$ in front of the exponential due to \cite{Massart}) and the fact that the Kolmogorov distance upper bounds $ \frac{1}{2} \ell_\infty $ distance (see, e.g., \cite[Section 4]{Canonne}). 

\begin{theorem}
\label{thm:learning-distr-linfty}
Let $P$ be a distribution on $\bR$.
Let $ \bfx_1,\cdots,\bfx_m $ be i.i.d.\ according to $P$ and denote by $ {P}_m $ their empirical distribution:
\begin{align}
{P}_m(x) &= \frac{1}{m}\sum_{i = 1}^m\indicator{\bfx_i = x}, \notag 
\end{align}
for any $ x\in\bR $.
Then for every $\eps>0$,
\begin{align}
\prob{\norm*{{P}_m - P}_\infty\ge\eps} 
\le 2\cdot \exp\paren*{-\frac{1}{2}m\eps^2}. \notag 
\end{align}
\end{theorem}

\section{A lemma regarding performance of random hash functions}

For each seed $\vr\in\Brackets{q}^{\sql}$ draw, independently accross $\vr$'s, a hash table
\begin{equation*}
    h(\cdot,\vr):\ \Brackets{q}^{\ell}\longrightarrow \Brackets{q}^{\,\ell H_q(\varrho)(1+\delta)}
\end{equation*}
uniformly at random. Given $\vx,\vs\in\Brackets{q}^{\ell}$ with $\wth{\vs}\le \ell \varrho(1+\eps)$, let $\vy=\vx+\vs$ and call a seed $\vr$ \emph{bad} if
\begin{equation*}
    \exists\,\tvx\in\cB\Bigl(\vy,\ell \varrho(1+\eps)\Bigr)\setminus\{\vx\}
    \quad\text{s.t.}\quad
    h(\tvx,\vr)=h(\vx,\vr).
\end{equation*}
Let
\begin{equation*}
    B(\vx,\vs)\;=\;\bigl|\{\,\vr\in\Brackets{q}^{\sql}:\ \vr\ \text{is bad}\,\}\bigr|
\end{equation*}
be the number of bad seeds for the fixed pair $(\vx,\vs)$. All probabilities below are over the draw of $\{h(\cdot,\vr)\}_r$.

\begin{lemma}[Few bad seeds]\label{lem:few-bad-seeds}
For all sufficiently large $\ell$,
\begin{equation*}
\Pr\parens*{B(\vx,\vs)\ge q^{\sqrt{\ell}/2}}
\;\le\;
\exp\Bigl(-\tfrac{1}{3}\,q^{\sqrt{\ell}/2}\Bigr).
\end{equation*}
\end{lemma}
\begin{proof}
Let $\cB=\cB(\vy,\ell \varrho(1+\eps))$. For a fixed seed $\vr$ and any $\tvx\neq\vx$,
\begin{equation*}
    \Pr\parens*{h(\tvx,\vr)=h(\vx,\vr)}=q^{-\ell H_q(\varrho)(1+\delta)}.
\end{equation*}
Union bound over $\tvx\in \cB\setminus\{\vx\}$ with $|\cB| \le q^{\ell\,H_q(\varrho)(1+\eps)}$, so the per-seed bad-event probability satisfies
\begin{align*}
\Pr\parens*{r\ \text{is bad}}
&\le
q^{\ell H_q(\varrho)(1+\eps)}\,q^{-\ell H_q(\varrho)(1+\delta)} \\
&=
q^{-\ell H_q(\varrho)(\delta-\eps)}.
\end{align*}
Let $E_{\vx,\vs}(\vr)$ be the event that seed $\vr$ is bad. Because the tables $\{h(\cdot,\vr)\}$ are drawn independently across $\vr$'s, the variables $\{\indicator{E_{\vx,\vs}(\vr)}\}$ are i.i.d. Bernoulli with
\begin{equation*}
\expt{\indicator{E_{\vx,\vs}(\vr)}}\;\le\;q^{-\ell H_q(\varrho)(\delta-\eps)}.
\end{equation*}
Hence
\begin{align*}
B(\vx,\vs)&=\sum_{r\in\Brackets{q}^{\sql}} \indicator{E_{\vx,\vs}(\vr)} \\
\mu&:=\expt{B(\vx,\vs)}
\le
q^{\sql}\,q^{-\ell H_q(\varrho)(\delta-\eps)} \\
&\hspace{62pt}= q^{\,\sql-\ell H_q(\varrho)(\delta-\eps)}.
\end{align*}
The multiplicative Chernoff bound yields
\begin{equation*}
\Pr\parens*{B(\vx,\vs)\ge q^{\sqrt{\ell}/2}}
\;\le\;
\exp\Bigl(-q^{\sqrt{\ell}/2}\cdot\sqrt{\ell}\cdot(\delta-\epsilon)\Bigr). \qedhere
\end{equation*}
\end{proof}
\bibliographystyle{alpha}
\bibliography{ref} 

\end{document}